\documentclass[fleqn,usenatbib]{mnras}

\usepackage{newtxtext,newtxmath}

\usepackage{subcaption}

\usepackage[T1]{fontenc}

\DeclareRobustCommand{\VAN}[3]{#2}
\let\VANthebibliography\thebibliography
\def\thebibliography{\DeclareRobustCommand{\VAN}[3]{##3}\VANthebibliography}

\usepackage{graphicx}	
\usepackage{amsmath}	
\usepackage{multirow}
\usepackage{rotating}
\usepackage{xcolor}

\graphicspath{./figures/}

\newenvironment{referee_comments}{\color{black}}

\title{Dust Morphology Under Changing Dust Mass Ratios in Protoplanetary Discs}
\author[M. H. Murray et al.]{
Matthew H. Murray,$^{1,2}$\thanks{E-mail: matthew.murray@uga.edu}
C. Hall,$^{1,2}$,
Hans Baehr,$^{1,2}$,
Jason P. Terry,$^{1,2,3}$\\
$^{1}$Department of Physics and Astronomy, The University of Georgia, Athens, GA 30605\\
$^{2}$Center for Simulational Physics, The University of Georgia, Athens, GA 30605\\
$^{3}$Department of Earth Sciences, University of Oxford, Oxford, UK}

\date{Accepted XXX. Received YYY; in original form ZZZ}

\pubyear{2025}

\begin{document}
\label{firstpage}
\pagerange{\pageref{firstpage}--\pageref{lastpage}}
\maketitle

\begin{abstract} 
Protoplanetary disc mass is one of the most fundamental properties of a planet-forming system, as it sets the total mass budget available for planet formation. However, obtaining disc mass measurements remain challenging, since it is not possible to directly detect H$_2$, and CO abundance ratios are poorly constrained. Dynamical measurements of the disc mass are now possible, but they are not suited to all discs since the measurements typically require well-behaved emission surfaces. A long-standing method is to obtain continuum flux measurements from the dust emission, and convert to a total disc mass by assumption of the dust-to-gas mass ratio, $\epsilon$. This quantity is poorly constrained in protoplanetary discs. 
We investigate the impact of $\epsilon$ on the morphology of planet-containing hydrodynamical simulations of dusty protoplanetary accretion discs, and suggest that if a planet mass estimate can be obtained, then disc morphology could be used to constrain $\epsilon$ in observed systems relative to each other, improving the total disc mass estimates of protoplanetary discs. 

\end{abstract}





\section{Introduction}

There are now over 5000 confirmed exoplanet detections, revealing huge diversity in the exoplanet architecture. Arguably the most fundamental exoplanet properties are semi-major axis and mass, which vary over four to five orders of magnitude in the known population \citep{zhudong2021}. A planet's semi-major axis may be heavily influenced by planet-disc interactions that lead to migration early in the disc's life \citep{Kley2012}, but may also continue to evolve under secular interactions well after disc dispersal \citep{ida_orbital_2000}. Similarly, the exoplanet mass is determined by disc properties, as more massive discs generally have more material available for planet formation, and may contain a higher density of solid material, particularly in the disc innermost regions. Obtaining an accurate measurement of protoplanetary disc (PPD) mass is therefore crucial to understanding the diversity of exoplanets.

Disc mass is also one of the most difficult characteristics to constrain through observations. PPDs are mostly made of molecular hydrogen (H$_2$), which emits very weakly at typical disc temperatures \citep{carmona_observational_2010}.  As a result, total mass estimates are challenging to obtain directly, and are typically inferred indirectly.

One of the most common methods of determining PPD mass involves using a tracer molecule such as CO, which is $\sim$ 10 million times more emissive than H$_2$ at typical disc temperatures \citep{bergin2017determination}. However, this is not without complications. The lines of some CO isotopologues may be optically thick, therefore only tracing the upper layers of the disc and masking emission from deeper layers. This scenario gives incomplete account of the total emission from that tracer. Additionally, there are uncertainties in the assumed abundance ratios which may be affected and altered by chemical reactions and photodissociation. An alternative that has emerged recently is the measurement or constraint of disc mass through dynamical \citep{veronesi2021,lodato2023,veronesi2024} or kinematic means \citep{hall2020,longarini2021,terry2022,pinte2023,speedie2024}, although this requires that the disc be sufficiently massive to be self-gravitating.

Historically, one of the most common methods to obtain disc mass has been to use the continuum emission to obtain the total dust mass \citep{andrews2012}, and then infer both the gas and total disc mass based on an assumed dust-to-gas ratio, $\epsilon$. However, there are significant sources of uncertainty. For example, dust mass is inferred based on assumptions about dust opacity, an unknown quantity that can vary widely depending on temperature, composition, size, and porosity. This can lead to incorrect inference about whether the emission is optically thin or optically thick at the observed wavelengths.\citep{xin_measuring_2023}.

Perhaps the strongest source of uncertainty is the assumed $\epsilon$. Best estimates for $\epsilon$ in the interstellar medium (ISM) are $\epsilon \sim 0.01$\citep{frisch_chemical_2003,frisch_interstellar_2011,nguyen_dustgas_2018}, and as a result this is a commonly assumed value for PPDs. However, this is likely an oversimplification. For example, observations of CO have found that the average disc $\epsilon$ values can be substantially higher, reaching around 0.2 \citep{ansdell2016}. Similarly, simulations, such as those by \citet{Lebreuilly2020}, have shown that discs can form with significant dust enrichment. Compounding this is the large discrepancy between the observed radial extent of the gas and dust disc components, with the latter typically appearing far more compact \citep{andrews2012,perez2012,perez2015} due to radial drift of the dust grains \citep{weidenschilling1977}. In the most extreme cases, such as IM Lup \citep{cleeves2016}, the dust disc can be 10 times more compact than its gas component, and since $\epsilon = \epsilon_{\mathrm{ISM}}[{R_{\text{gas disc}}}  /{R_{\text{dust disc }}}]^2$ \citep{ileehall2020}, this could potentially result in $\epsilon$ values as high as $\epsilon=1.0$ in the inner $\sim$ 120 au of the disc.

At the other end of the scale, the exoALMA sample has demonstrated that the dust-to-gas ratio may be as extreme as 1:400 in some systems \citep{longariniexoALMA}, suggesting depletion relative to the ISM.

Observations with the Atacama Large Millimeter/submillimeter Array (ALMA) over the last decade have identified a plethora of 
substructure within PPDs, such as rings or gaps \citep{alma_partnership_2014_2015,andrews_disk_2018} and spirals \citep{perez2016,huang_disk_2018}. Thanks to the discovery of velocity perturbations in CO line emission \citep{Pinte2018,Teague2018,paneque2021,speedie2024}, the origin of these substructures is now becoming clear. In the case of continuum rings and gaps, the presence of localised velocity perturbations is leading to a growing consensus that the gaps are caused by forming protoplanets. While there are other plausible origins (e.g. snow lines), it has been shown that in general the radial distance of the gaps from the central star are inconsistent with snow lines of common molecules \citep{marel2019}, and furthermore the snow lines cannot explain the numerous kinematic features now known \citep[e.g.][]{pinte_kinematic_2019,pinte_nine_2020,bae_molecules_2022,terry2023}

When discussing gaps induced by planetary action, morphology is determined by a combination of disc and planet properties, with prescriptions that have been determined both analytically and empirically from numerical simulations \citep[see, e.g.][]{crida_width_2006,kanagawa_mass_2015,kanagawa_formation_2015,kanagawa_mass_2016,tanaka_eccentric_2022}. 
Of particular interest here is the morphology of the dust gap, which can be observed at $\sim$mm wavelengths with the ALMA telescope. The \textit{width} of the dust gap can be described purely analytically in terms of disc parameters and dust properties such as the dust-to-gas ratio $\epsilon$ \citep{dipierro_opening_2017}, but there is currently no analytical prescription for dust gap \textit{depth}. 
Additionally, the modification of the gas surface density due to the dust back reaction can result in a feedback loop between dust and gas that further alters dust surface density and morphology. 
The effects of the back reaction are non-linear, so they are best addressed numerically using simulations.

In this work, we demonstrate how changing the global dust-to-gas ratio, $\epsilon$, affects the dust gap depth of a PPD with an embedded planet. We show that this could potentially be used to constrain the $\epsilon$ values in observed systems, in particular in systems where protoplanet mass is already constrained by disc kinematics or other means. 
Additionally, we show that the current analytical prescription for gap depth fails to accurately predict how both the \textit{gas} gap depth and the \textit{dust} gap depth change with respect to $\epsilon$. We take this as further indication of the importance of the back reaction when simulating PPDs.

The paper is organized as follows: in Section \ref{Methods},  we describe our numerical simulations, and discuss our choice of parameter space and measurement of morphological properties. In Section \ref{Results}, we describe our results, and discuss the different scenarios arising from our choice of initial conditions and the physics we include.  In Section \ref{Conclusion}, we present our conclusions and discuss the limitations of our work, as well as opportunities for future work and potential questions raised by this study. 

\section{Method}\label{Methods}

\subsection{Smoothed Particle Hydrodynamics}\label{SPHModel}

We perform nineteen three-dimensional hydrodynamical simulations of dusty, gaseous protoplanetary discs containing a single planet using \texttt{PHANTOM} \citep{price_phantom_2018}, a smoothed particle hydrodynamics (SPH) code 
\begin{referee_comments}
    that solves the equations of hydrodynamics in Lagrangian form \citep{lucy1977,gingold_smoothed_1977} by discretising the fluid onto a set of particles. Full algorithm details are given in \citet{price_phantom_2018} and references therein, but in summary the equations of compressible hydrodynamics are solved in the form:
\begin{equation}
\begin{aligned}
\frac{\mathrm{d} \boldsymbol{v}}{\mathrm{~d} t}= & -\frac{\nabla P}{\rho}+\Pi_{\text {shock }}+\boldsymbol{a}_{\text {ext }}(\boldsymbol{r}, t) +\boldsymbol{a}_{\text {sink-gas }}+\boldsymbol{a}_{\text {selfgrav }}
\end{aligned}
\end{equation}

\begin{equation}
\frac{\mathrm{d} u}{\mathrm{~d} t}=-\frac{P}{\rho}(\nabla \cdot \boldsymbol{v})+\Lambda_{\text {shock }}-\frac{\Lambda_{\text {cool }}}{\rho},
\end{equation}

\begin{referee_comments}
    where $P$ is the pressure, $u$ is the specific internal energy, $\boldsymbol{a}_{\text {ext }}, \boldsymbol{a}_{\text {sink-gas }}$ and $\boldsymbol{a}_{\text {selfgrav }}$ refer to external, sink and self-gravity forces present in the system. $\Pi_{\text {shock }}$ and $\Lambda_{\text {shock }}$ are dissipative terms required to calculate entropy correctly at a shock front, and $\Lambda_{\text {cool }}$ is a cooling term.
\end{referee_comments}

\end{referee_comments}

\begin{referee_comments}
    Fluid elements are discretized when calculating field terms and density, $\rho(\textbf{r)}$, is obtained from discrete fluid elements using a smoothing kernel \citep{gingold_smoothed_1977}:
    \begin{equation}
        \rho_\mathrm{s}(\boldsymbol{r})=\int W(\boldsymbol{r-\boldsymbol{r}'})\rho(\boldsymbol{r}')d\boldsymbol{r}'
    \end{equation}
Where $W$ satisfies: 
\begin{equation}
            \int W(\boldsymbol{r})d\boldsymbol{r} = 1
\end{equation}
The choice of the smoothing kernel and smoothing length, $h$, may vary with respect to the given problem. The default smoothing kernel in \texttt{PHANTOM} is an $M_{4}$ cubic spline, the SPH standard, and $h$ is calculated on each step to preserve local density \citep{monaghan_refined_1985, price_phantom_2018}
\end{referee_comments}

We use the one-fluid dust method for its ability to handle small grains and low Stokes numbers \citep{laibe_dusty_2014, laibe_dusty_2014-1,laibe_dust_2014,hutchison2018,ballabio2018enforcing}, which follows the dust fraction on each particle. 
\begin{referee_comments}
    Under this prescription there is a single SPH "fluid" moving with respect to the centre of mass. The barycentric velocity is then: 
    \begin{equation}
     \label{eq:barycentric_v}
         \boldsymbol{v} = \frac{\rho_\mathrm{g}\boldsymbol{v}_\mathrm{g}+\rho_\mathrm{d}\boldsymbol{v_\mathrm{d}}}{\rho_\mathrm{g}+\rho_\mathrm{d}}.
     \end{equation}
    Where $\Delta\boldsymbol{v}=\boldsymbol{v_\mathrm{g}}-\boldsymbol{v_\mathrm{d}}$. 
    
    This produces a Lagrangian frame that is co-moving with the center of mass of the system\citep{laibe_dust_2014}. Leveraging the dust fraction, $\epsilon \equiv {\rho_\mathrm{d}}{\rho^{-1}}$, it is also possible to remove singularities that may arise from especially small $\rho_\mathrm{g}$, while also providing direct leverage on a quantity of interest.
\end{referee_comments}

We used 1 million SPH particles, and simulated the central star as a sink particle \citep{bate_modelling_1995} of $1 \text{M}_{\odot}$ and accretion radius of 0.8 au. The embedded planet is also modelled as a sink particle, with a mass of $2 \text{M}_{\text{J}}$ and an accretion radius is set to $0.25$ of the Hill radius for the planet \citep{hill_researches_1878}. The planet's initial orbital radius is set to 60 au, and is not held on a fixed orbit. All discs are initialized with an inner radius of 1 au and an outer radius of 120 au. Surface density and sound speed profiles follow $\Sigma \propto R^{-1}$ and $c_\mathrm{s} \propto R^{-0.25}$ respectively. $\Sigma_0$ is set by the disc masses described in Tables \ref{tab:MassTableGrainBRon}, \ref{tab:MassTableGrainBRoff}, \ref{tab:MassTableStokesBRon}, and \ref{tab:MassTableStokesBRoff}. Discs are evolved for 80 orbits at the location of the planet,  
\begin{referee_comments}
     which is comparable to existing work and benchmarks for timescales on which gaps in dusty discs are able to form \citep{dipierro_two_2016, pinte_kinematic_2023, bae_exoalma_2025}.
\end{referee_comments}

The $\alpha_\mathrm{SPH}$ value is set in each disc such that the Shakura-Sunyaev artificial viscosity parameter, $\alpha_\mathrm{SS}$, is  $\alpha_\mathrm{SS} \approx 0.005$ \citep{ShakuraSunyaev1973}. This is consistent with values obtained from observations, which typically obey  $10^{-4} \lesssim \alpha \lesssim 10^{-2}$(\citealt{MulderDominikVisc,Pinte2016,zhang_disk_2018}). For all discs, we set $\beta_\mathrm{SPH}=2.00$. 
\begin{referee_comments}
    This is the quadratic Von Neumann-Richtmyer artificial viscosity term \citep{ging}(Gingold/Monaghan 1983), controlling shock dissipation, particle overlap, and particle oscillations that cross the disc's midplane.
\end{referee_comments}

We investigate two scenarios: The first is dust of a constant grain size, $a =0.1$ mm, with the size chosen as the typical grain size that is probed through $\sim$mm observations with the ALMA telescope (hereafter ``constant grain size'').

The second scenario is dust of a constant Stokes number, St (hereafter ``constant St''), which is defined as the ratio between the stopping (or response) time and the dynamical time (or characteristic timescale of the system) such that
$\mathrm{St}=t_{\mathrm{s}}/t_{\mathrm{dyn}}$. For $\mathrm{St}\ll 1$, particles are perfectly coupled to the gas since the response time is small. For $\mathrm{St}\gg 1$, particles are decoupled, and do not respond (or respond slowly) to changes in the fluid flow. At St$\sim$1, both the dust and the fluid have similar reaction times to changes in the local conditions, which results in the dust both feeling the gas drag and having independent motion. As a consequence, St = 1 particles feel the largest headwind, and therefore have the most rapid inward motion \citep{weidenschilling1977,whipple1972certain}. In a PPD, St can be expressed as \citep{Birnsteil2010}:
\begin{equation}
    \mathrm{St}=\frac{\pi a \rho_{\mathrm{s}}}{2 \Sigma_{\mathrm{g}}}
\end{equation}
where $\rho_\mathrm{s}$ is grain density and $\Sigma_\mathrm{g}$ is the two-dimensional gas surface density. We perform simulations with a constant value of St to keep the aerodynamic behaviour of the grains constant throughout the simulation. A value of St = 0.1 is chosen as it is close enough to St = 1.0 to capture the most interesting grain behaviour while being computationally efficient enough to allow the simulation to proceed without encountering timestepping problems.

We implement a constant St in \texttt{PHANTOM} through the stopping time according to:   
\begin{equation}
    t_\mathrm{s} = \frac{\mathrm{St}}{\Omega_\mathrm{k}},
\end{equation}
where $\Omega_\mathrm{k}$ is the Keplerian angular frequency and is equal to $1/t_\mathrm{dyn}$.

In both scenarios, we vary $\epsilon$ as well as the back reaction of the dust onto the gas, turning it on and off the to explore its effect on disc morphology. 

For a given annulus, the relative azimuthal velocity of the gas and the dust imparts a drag torque, $\Lambda_{\mathrm{g}\rightarrow \mathrm{d}}$, from the gas to the dust of the form \citep{dipierro_opening_2017}:
\begin{equation}
    \label{eq:drag1}
         \Lambda_{\mathrm{g}\rightarrow \mathrm{d}}= -r\frac{K}{\rho_\mathrm{d}}(v_{\mathrm{d},\theta}-v_{\mathrm{g},\theta}),
\end{equation}
and the back reaction is an opposite torque, with magnitude scaled by $\epsilon$:
\begin{equation}
    \label{eq:BackRe1}
        \Lambda_{\mathrm{d}\rightarrow \mathrm{g}} = -\epsilon \Lambda_{\mathrm{g}\rightarrow \mathrm{d}}.
\end{equation}
The constant $K$ is the drag coefficient \citep{LaibePrice_2012}, and is a function of the stopping time, $t_\mathrm{s}$, and the dust-to-gas mass ratio, $\epsilon$:
\begin{equation}
    K = \frac{\rho_\mathrm{d}}{t_\mathrm{s} + \epsilon}.
\end{equation}
\begin{referee_comments}
    In \texttt{PHANTOM}, when the back reaction is "on", $\Lambda_{\mathrm{g}\rightarrow \mathrm{d}}$ is calculated following Equation \ref{eq:drag1} and $\Lambda_{\mathrm{d}\rightarrow \mathrm{g}}$ is found based on local gas and dust densities (Equations 239 and 240 in \cite{price_phantom_2018}). This allows the back reaction to vary as $\epsilon$ changes throughout a disk. When the back reaction is "off", \texttt{PHANTOM} skips calculating $\Lambda_{\mathrm{d}\rightarrow \mathrm{g}}$ entirely, and applies no drag to the gas phase. 
\end{referee_comments} 
When considering the back reaction, net torque on a given annulus of the disc decreases linearly with $\epsilon$:
\begin{align}\label{eq:NetTorque}
    \Lambda_\mathrm{net} &= \Lambda_{\mathrm{g}\rightarrow \mathrm{d}} + \Lambda_{\mathrm{d}\rightarrow \mathrm{g}}\notag\\
     &= \Lambda_{\mathrm{g}\rightarrow \mathrm{d}} - \epsilon\Lambda_{\mathrm{g}\rightarrow \mathrm{d}}\notag\\
     &= \Lambda_{\mathrm{g}\rightarrow \mathrm{d}}(1-\epsilon).
\end{align}
To explore how this changes the disc morphology, we perform simulations at five different values of $\epsilon$: $0.005$, $0.01$, $0.05$, $0.1$ and $0.5$. 
For the constant grain size scenario that includes the back reaction, we hold total dust mass ($M_\mathrm{dust}$) constant at $1 \times 10^{-4}$ M$_{\odot}$ and vary $M_\mathrm{gas}$. 

The reason we do this is that mass estimates from continuum fluxes actually probe the amount of dust in a system, $M_\mathrm{dust}$ not total disc mass, $M_\mathrm{T}$. Instead, total disc mass is inferred from an assumed $\epsilon$ - an unknown quantity that may be depleted or enhanced in the disc relative to the local ISM \citep{ileehall2020,longariniexoALMA}. With this parameter space we are posing the question: for a given dust mass, how would the dust morphology observed at $\sim$mm wavelengths change for a varying $\epsilon$ and gas mass? The natural question that follows this is: can this change in disc morphology be used to constrain $\epsilon$, and therefore the total disc mass? If multiple systems have similar observed $M_\mathrm{dust}$ but different morphologies, this framework could give us a qualitative stepping stone to obtaining $M_\mathrm{T}$.

Simulation gas masses, dust masses, total masses and $\epsilon$  are given in Tables \ref{tab:MassTableGrainBRon} and \ref{tab:MassTableGrainBRoff} for the constant grain size scenarios, and Tables \ref{tab:MassTableStokesBRon} and \ref{tab:MassTableStokesBRoff} for the constant St scenarios. In simulations of constant grain size without the back reaction, total gas mass ($M_\mathrm{gas}$) is held constant at $1\times 10^{-3}$ $\text{M}_{\odot}$, and vary $M_\mathrm{dust}$. For the constant $\mathrm{St}=0.1$ scenario, the total disc mass is held constant between simulations to explore similar physical regimes.

\begin{table}
    \centering
    \caption{Table of simulated disc masses for the constant grain size scenarios where the back reaction is on. Total dust mass is held constant and $\epsilon$ determines gas mass.}
    \begin{tabular}{l c|c|c|c}
        \hline
         $\epsilon$ & Dust Mass [$\text{M}_{\odot}$] & Gas Mass [$\text{M}_{\odot}$]& Total Mass [$\text{M}_{\odot}$]\\
         \hline \hline
         $0.005$ & $1.00\times 10^{-4}$ & $2.00\times 10^{-2}$ & $2.010\times 10^{-2}$ \\
         $0.01$  & $1.00\times 10^{-4}$ & $1.00\times 10^{-2}$ & $1.010\times 10^{-2}$ \\
         $0.05$  & $1.00\times 10^{-4}$ & $2.00\times 10^{-3}$ & $2.10\times 10^{-3}$ \\
         $0.1$   & $1.00\times 10^{-4}$ & $1.00\times 10^{-3}$ & $1.10\times 10^{-3}$ \\
         $0.5$   & $1.00\times 10^{-4}$ & $2.00\times 10^{-4}$ & $3.0\times 10^{-4}$ \\
    \hline
    \end{tabular}

    \label{tab:MassTableGrainBRon}
\end{table}

\begin{table}
    \centering
    \caption{Table of simulated disc masses for the constant grain size scenarios where the back reaction is off. Total gas mass is held constant and $\epsilon$ determines dust mass.}
    \begin{tabular}{l c|c|c|c}
        \hline
         $\epsilon$ & Dust Mass [$\text{M}_{\odot}$] & Gas Mass [$\text{M}_{\odot}$]& Total Mass [$\text{M}_{\odot}$]\\
         \hline \hline
         $0.005$ & $5.00\times 10^{-6}$ & $1.00\times 10^{-3}$ & $1.005\times 10^{-3}$ \\
         $0.01$  & $1.00\times 10^{-5}$ & $1.00\times 10^{-3}$ & $1.010\times 10^{-3}$ \\
         $0.05$  & $1.00\times 10^{-5}$ & $1.00\times 10^{-3}$ & $1.050\times 10^{-3}$ \\
         $0.1$   & $1.00\times 10^{-4}$ & $1.00\times 10^{-3}$ & $1.100\times 10^{-3}$ \\
         $0.5$   & $1.00\times 10^{-4}$ & $1.00\times 10^{-3}$ & $1.500\times 10^{-3}$ \\
    \hline
    \end{tabular}

    \label{tab:MassTableGrainBRoff}
\end{table}
\begin{table}
    \centering
    \caption{Table of simulated disc masses for the constant St numbers scenarios with the back reaction. Total gas mass is held constant with $\epsilon$ determining the dust mass budget. The $\epsilon=0.5$ simulation is included even though it is not used in our analysis due to being unable to run to the desired 80 orbits of the embedded planet.}
    \begin{tabular}{l c|c|c|c}
        \hline
         $\epsilon$ & Dust Mass [$\text{M}_{\odot}$] & Gas Mass [$\text{M}_{\odot}$]& Total Mass [$\text{M}_{\odot}$]\\
         \hline \hline
         $0.005$ & $5.000\times 10^{-6}$ & $1.005\times 10^{-3}$ & $1.010\times 10^{-3}$\\
         $0.01$  & $1.000\times 10^{-5}$ & $1.000\times 10^{-3}$ & $1.010\times 10^{-3}$\\
         $0.05$  & $5.000\times 10^{-5}$ & $9.500\times 10^{-4}$ & $1.010\times 10^{-3}$\\
         $0.1$   & $1.000\times 10^{-4}$ & $9.100\times 10^{-4}$ & $1.010\times 10^{-3}$\\
         $0.5$   & $5.000\times 10^{-4}$ & $5.100\times 10^{-4}$ & $1.010\times 10^{-3}$ \\
        \hline
    \end{tabular}
    \label{tab:MassTableStokesBRon}
\end{table}

\begin{table}
    \centering
    \caption{Table of simulated disc masses for the constant St numbers scenarios without the back reaction. Total gas mass is held constant with $\epsilon$ determining the dust mass budget.}
    \begin{tabular}{l c|c|c|c}
        \hline
         $\epsilon$ & Dust Mass [$\text{M}_{\odot}$] & Gas Mass [$\text{M}_{\odot}$]& Total Mass [$\text{M}_{\odot}$]\\
         \hline \hline
         $0.005$ & $5.050\times 10^{-6}$ & $1.00\times 10^{-3}$ & $1.010\times 10^{-3}$\\
         $0.01$  & $1.010\times 10^{-6}$ & $1.00\times 10^{-3}$ & $1.010\times 10^{-3}$\\
         $0.05$  & $5.050\times 10^{-5}$ & $1.00\times 10^{-3}$ & $1.010\times 10^{-3}$\\
         $0.1$   & $1.010\times 10^{-4}$ & $1.00\times 10^{-3}$ & $1.010\times 10^{-3}$\\
         $0.5$   & $5.050\times 10^{-4}$ & $1.00\times 10^{-3}$ & $1.010\times 10^{-3}$ \\
        \hline
    \end{tabular}
    \label{tab:MassTableStokesBRoff}
\end{table}
\subsection{Measurement of Planet Induced Dust Gap Depth}\label{Gaps}

Planet-induced gap morphology is a joint function of planet and disc properties that requires consideration of multiple parameters (see, e.g., 
\citealt{crida_width_2006,DuffelMac2013,fung_how_2014,kanagawa_formation_2015}). In this work, we measure planet-induced dust gap depth by adapting \citet{tanaka_eccentric_2022}, where the minimum surface density (deepest gap) in the gas phase of a disc is given by: 
\begin{equation}\label{eq:estimatedGap}
    \frac{\Sigma_\mathrm{gap}}{\Sigma_0} = \frac{1}{1+0.04K'}.
\end{equation}
Here $K'$ is defined as
\begin{equation}
\label{eq:KPrime}
 K' = \left(\frac{M_\mathrm{p}}{M_*}\right)^2 \left(\frac{h_\mathrm{p}}{r_\mathrm{p}}\right)^{-5}\alpha^{-1},
\end{equation}

where $h_\mathrm{p}$ is the disc scale height at the radial location of the planet, $r_\mathrm{p}$.

This analytical method for estimating gas gap depth and describing shape is widely used and accepted, however, it is likely not directly translatable to dust gap depth. Interplay between the back reaction of the dust onto the gas and vice-versa significantly complicates the dust dynamics with the gas. \begin{referee_comments}
    We use this gas gap depth method as an approximation, under the assumption that a well coupled dust disc will have a similar shape and profile to it's host gas disc. Additionally at the time of writing there is no widely accepted/used analytical method to capture the behaviour of specifically the dusty components of discs. 
\end{referee_comments}

To allow for consistent and direct comparison in simulations that allow $\epsilon$ to very over orders of magnitude, we reduce the azimuthally averaged disc profile following the method of \cite{crida_cavity_2007}. A logarithmic surface density is taken to be the  background surface density, and is subtracted from the azimuthally averaged surface density profile. We assume that an unperturbed disc has a settled dust surface density that follows the general form of a decaying exponential, given by: 
\begin{equation}\label{eq:ZETA}
    \Sigma_{\mathrm{background,dust}}= \Sigma_\mathrm{dust}\rvert_{R=20\mathrm{au}} R^{-\zeta},
\end{equation}
where $\Sigma_\mathrm{background,dust}$ is the fitted surface density profile that captures behaviour agnostic to the planet, $R$ is radius in au, $\Sigma_\mathrm{dust}\rvert_{R=20\mathrm{au}}$ is the reference surface density at $20$ au, and $\zeta$ is the fitting parameter. For a full report of $\zeta$ parameters found for each disc, see Appendix \ref{zetaGas}.

Once a background surface density profile, $\Sigma_\mathrm{background}$ is determined, we subtract that from the dust surface density profile of the given scenario:
\begin{equation}
\label{eq:deltasigma}
    \Delta\Sigma_\mathrm{dust} = \Sigma_\mathrm{simulated,dust} - \Sigma_\mathrm{background,dust}.
\end{equation}
Negative values of $\Delta\Sigma_\mathrm{dust}$ indicate gaps induced by planets, with the magnitude being the gap depth. We take two measurements for each simulated disc: the max gap depth over the whole profile, and the gap depth at the final planetary radius from the central star. We take two measurements because some of the azimuthally averaged dust profiles have a soft ``w'' shape, with the central peak of the ``w'' located at $r_p$ \citep{dipierroelias,meru_is_2018}. As a result, the deepest portion of the gap is not necessarily co-located with the planet.

\begin{figure*}
    \centering
    \begin{tabular}{cc}
        \includegraphics[width=0.5\textwidth]{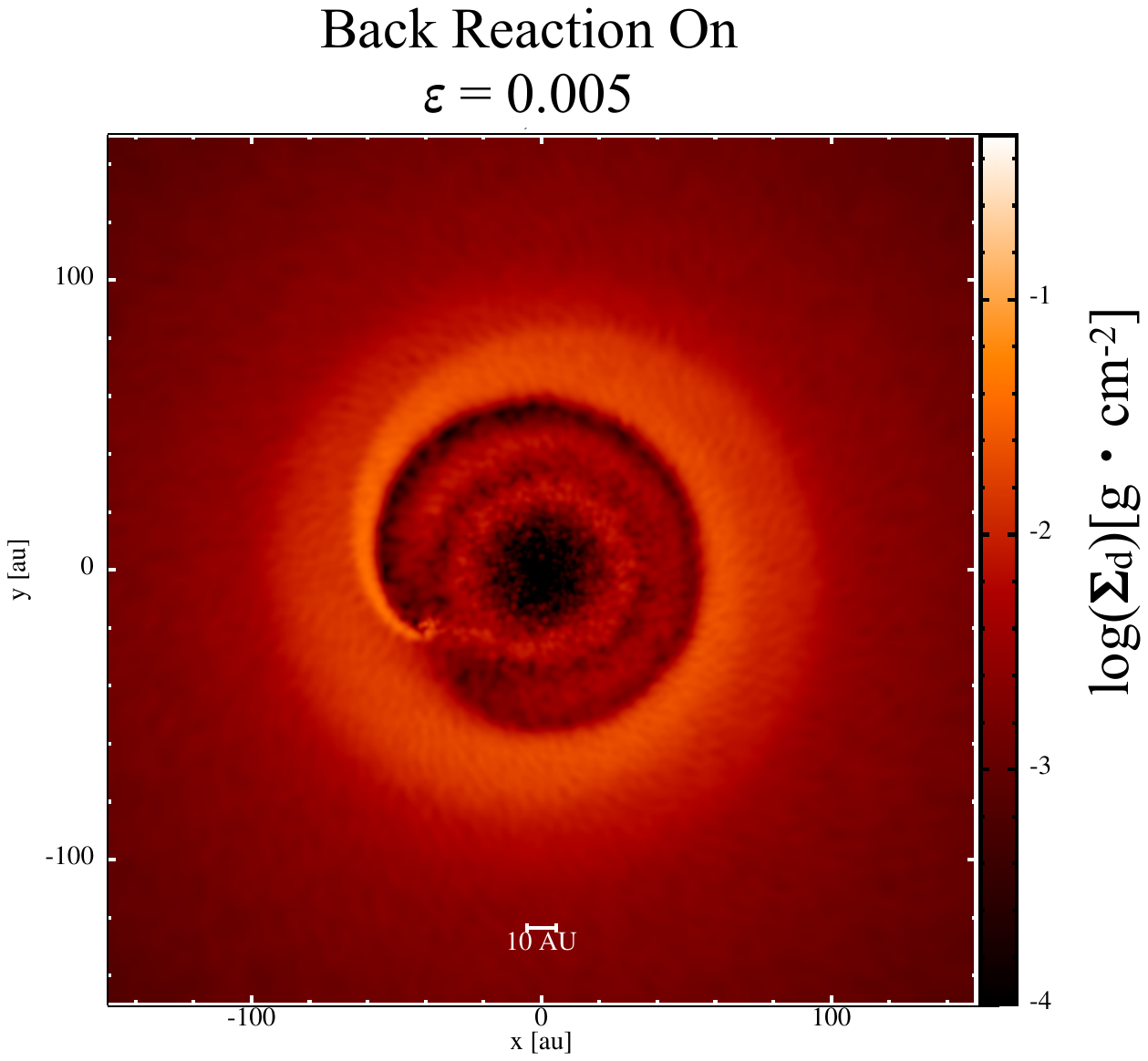} & \includegraphics[width=0.5\textwidth]{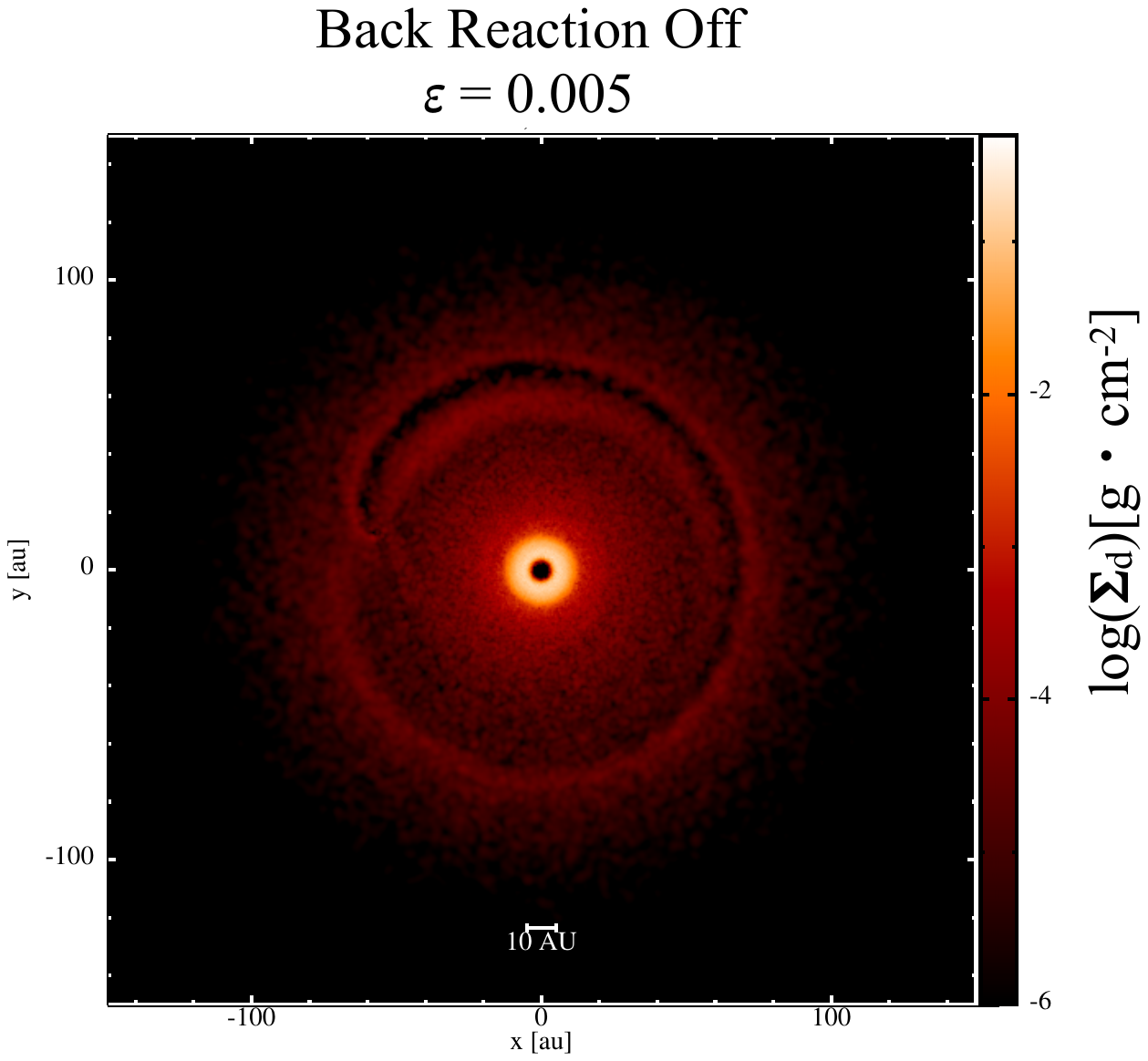} \\
        \includegraphics[width=0.5\textwidth]{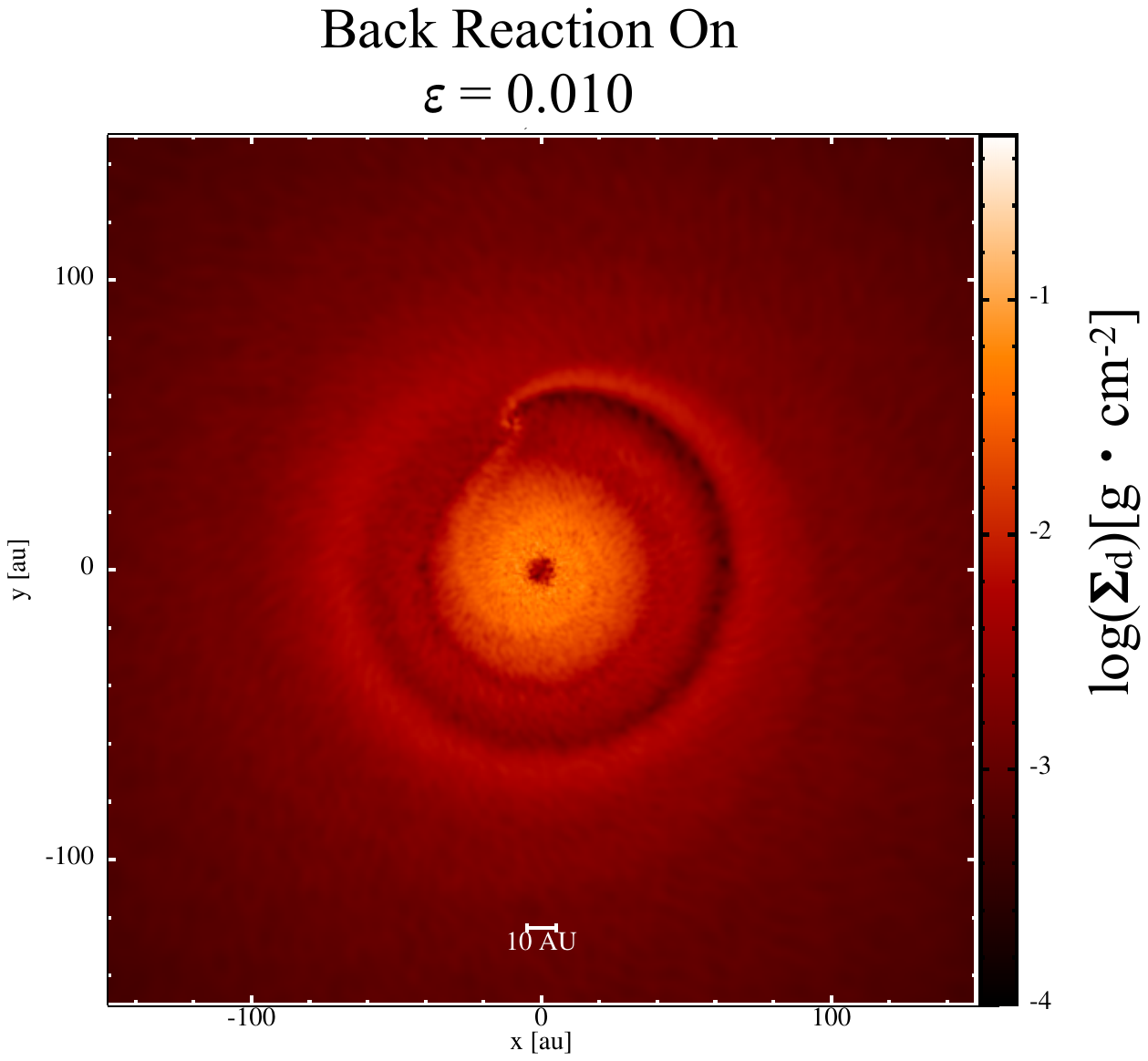} & \includegraphics[width=0.5\textwidth]{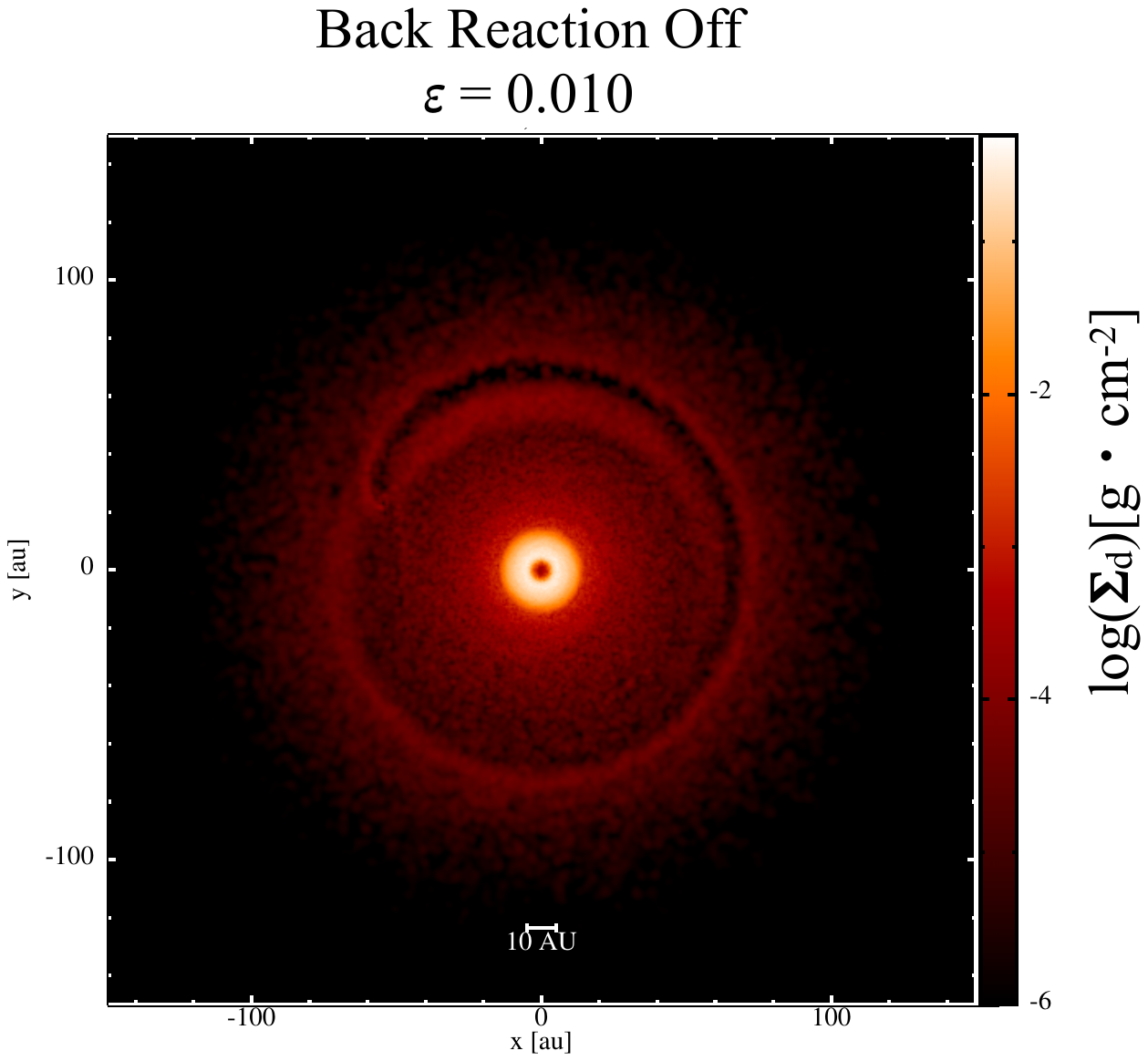} \\
        \includegraphics[width=0.5\textwidth]{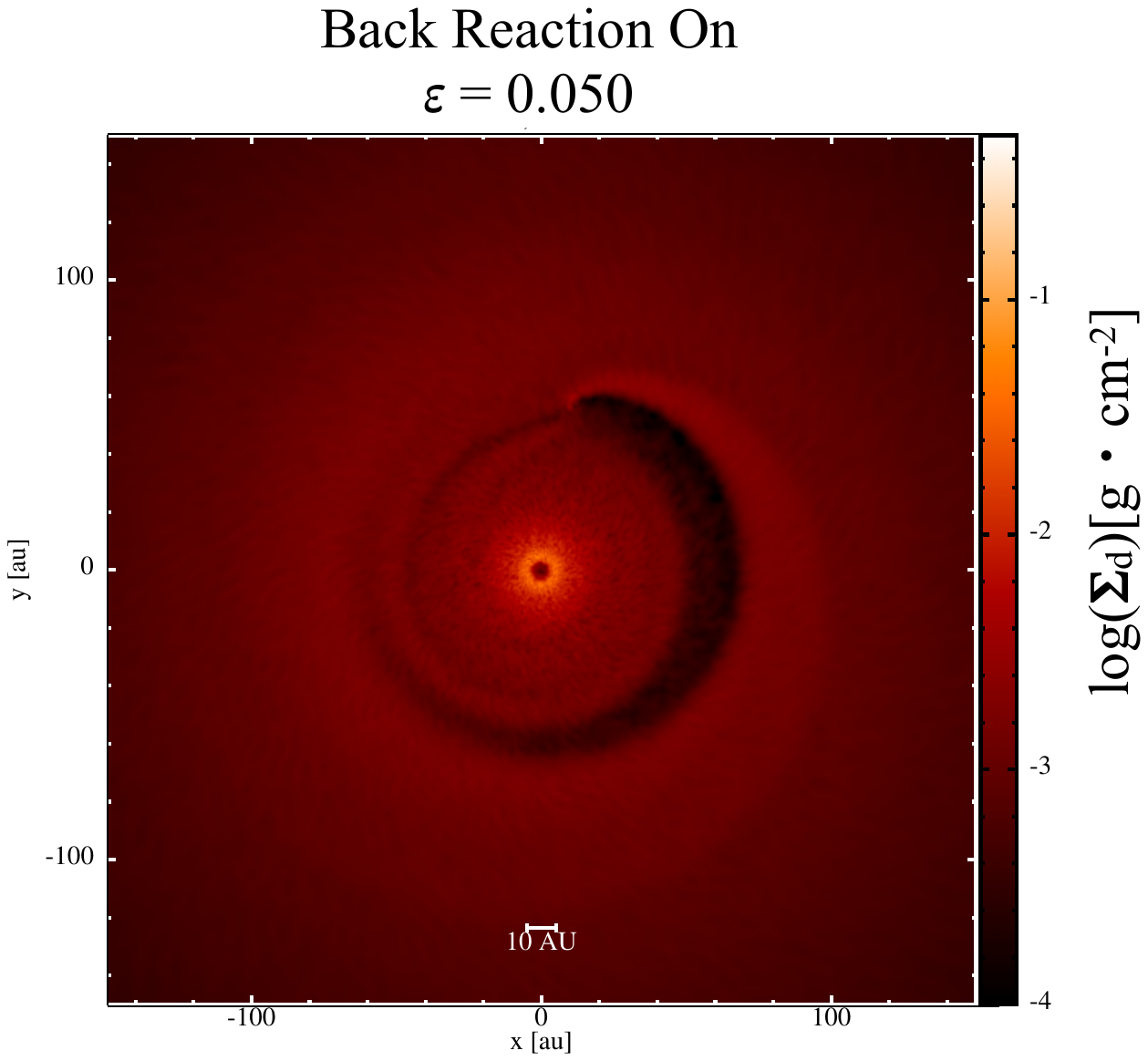} & \includegraphics[width=0.5\textwidth]{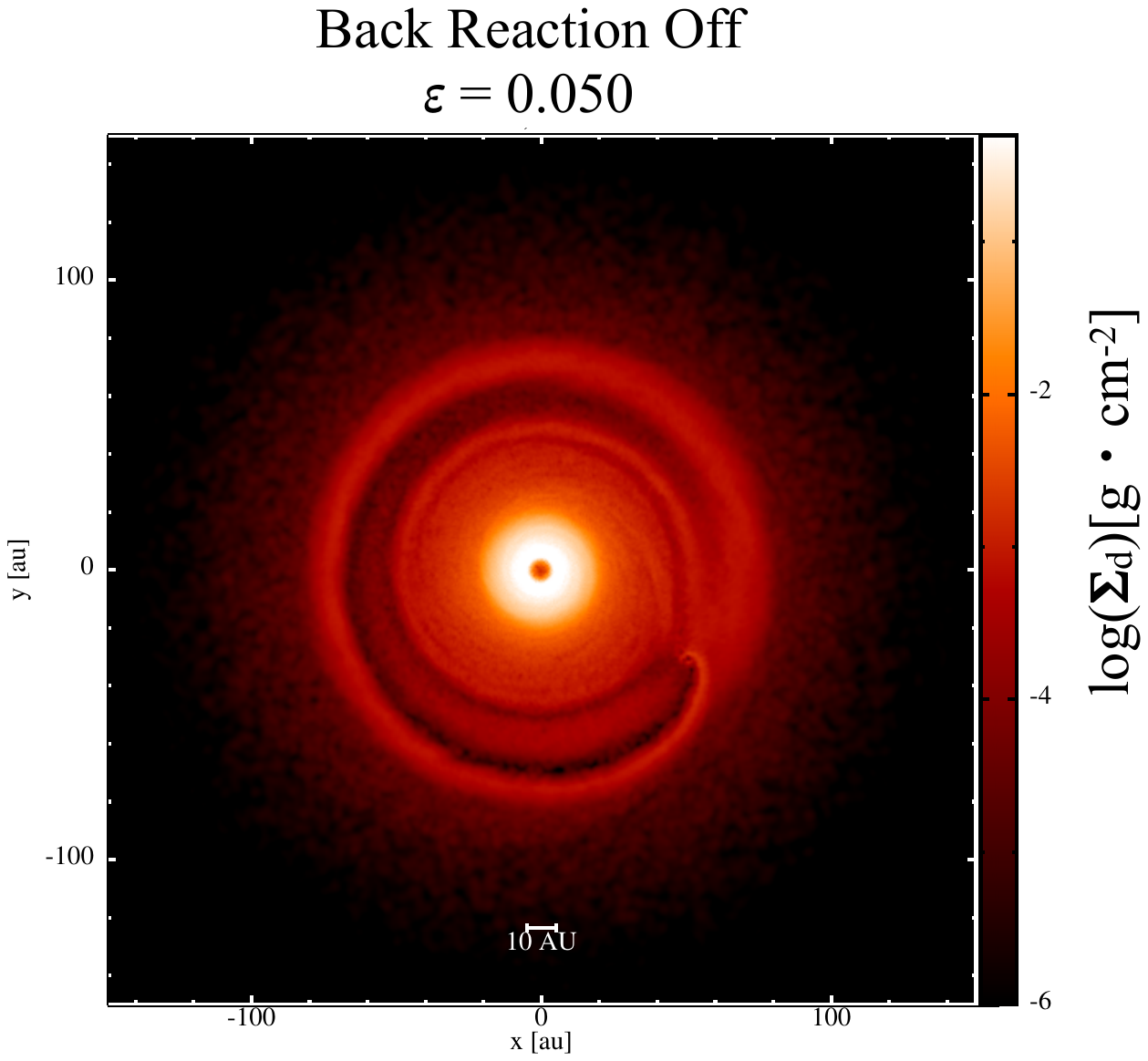} \\
    \end{tabular}
     \caption{Dust surface density of simulated discs with $a=0.1$mm. The left column is simulated with the back reaction, and the right column is simulated without the back reaction. Each row is a different $\epsilon$, increasing down the page and into Figure \ref{fig:Sarracen2}. At every simulated value of $\epsilon$, we see a remarkably different dust distribution. Discs without the back reaction have well defined outer edges at $\sim$100 au, and larger, higher density inner discs. Additionally, discs with and without the back reaction display different symmetries.}
     \label{fig:Sarracen1}
    \end{figure*}
    \begin{figure*}
    \begin{tabular}{cc}
        \includegraphics[width=0.5\textwidth]{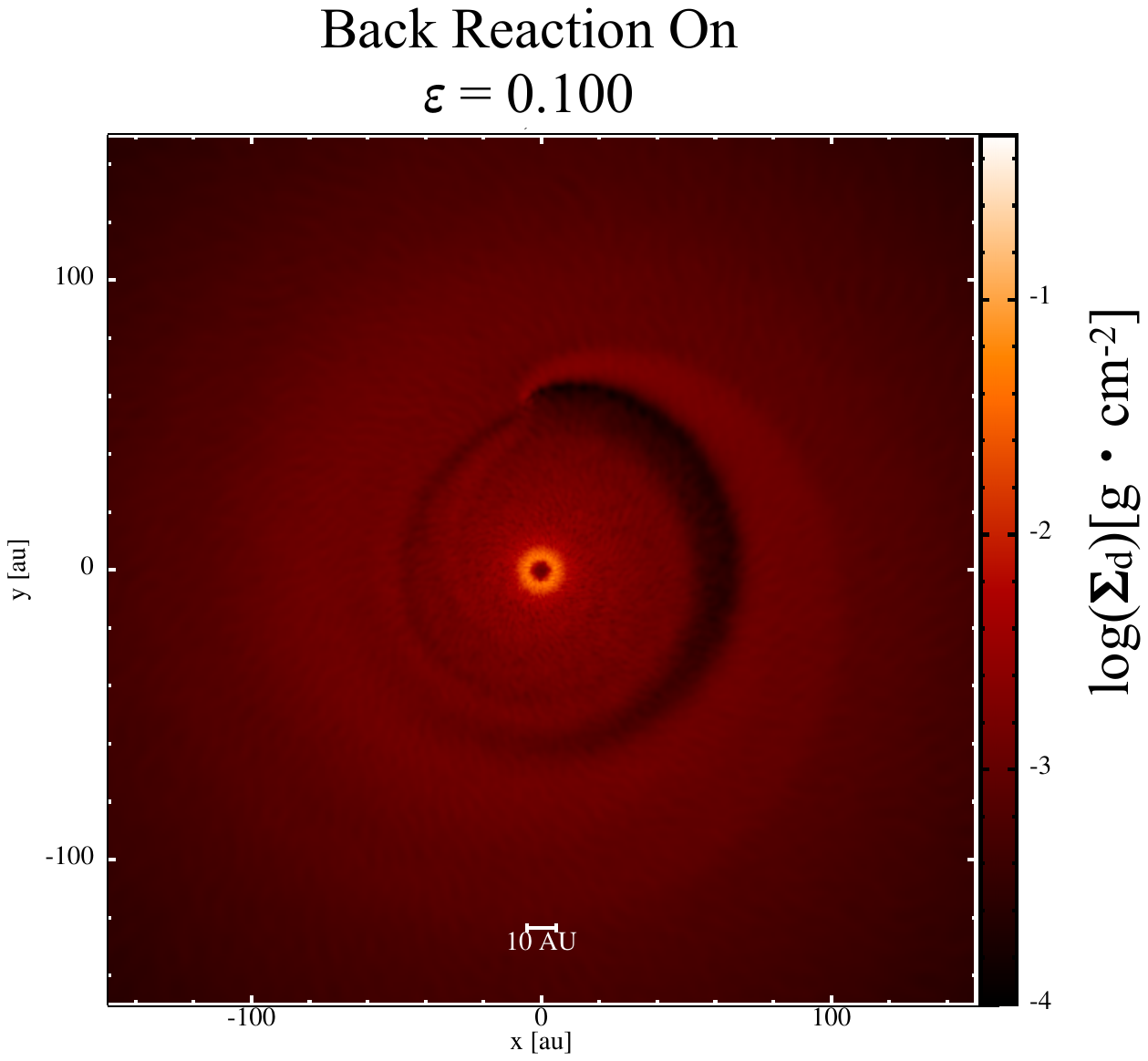} & \includegraphics[width=0.5\textwidth]{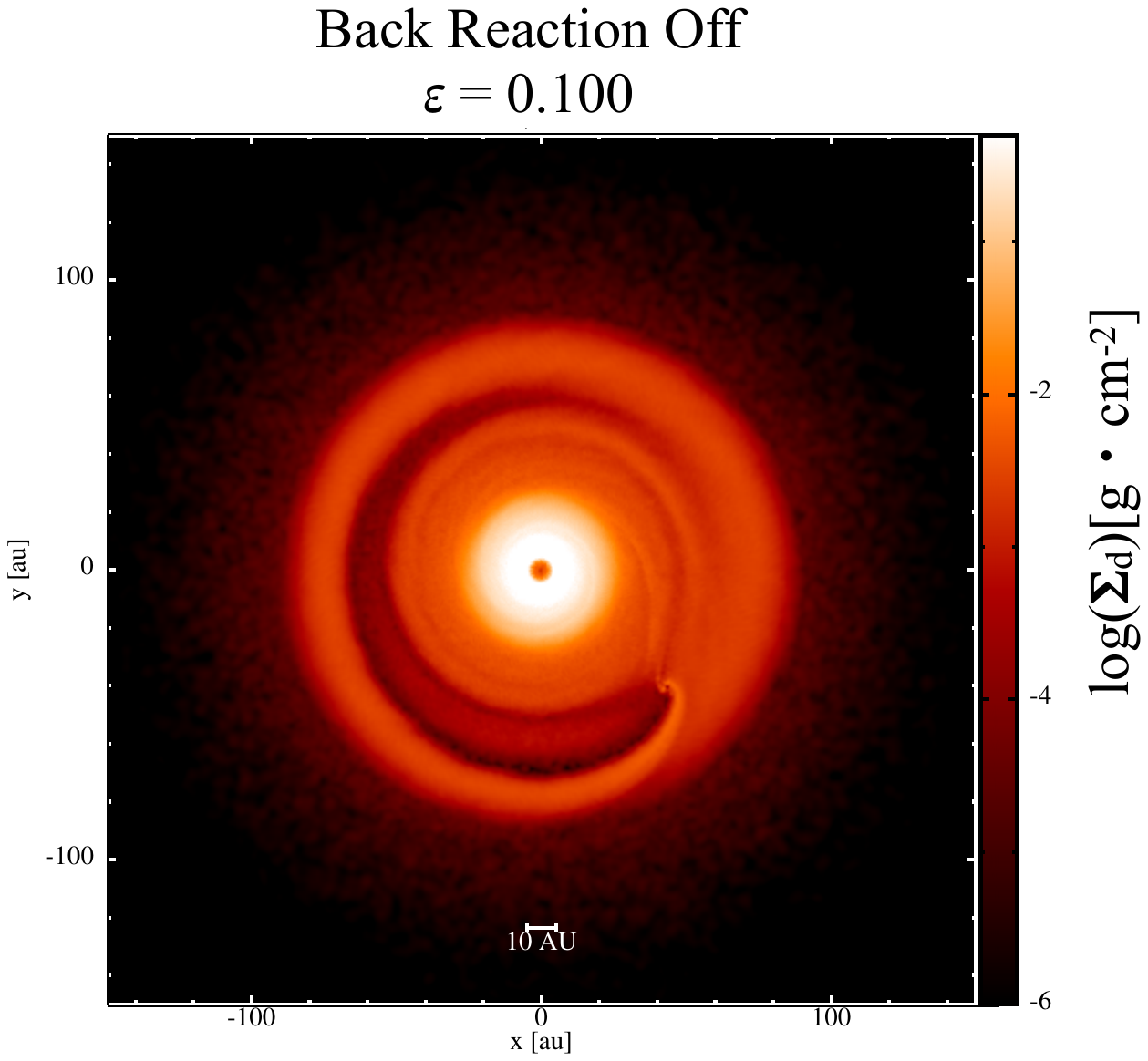} \\
        \includegraphics[width=0.5\textwidth]{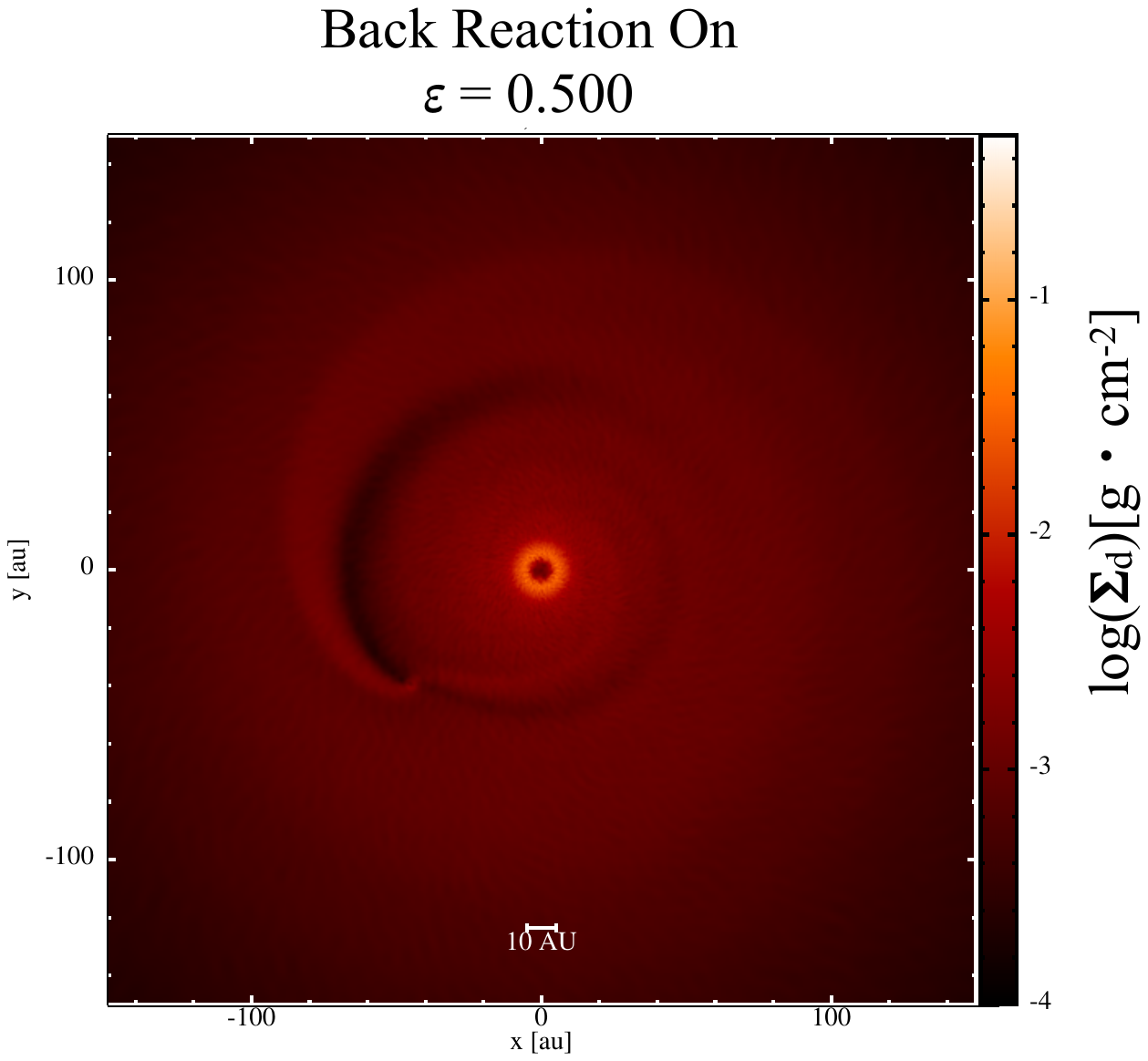} & \includegraphics[width=0.5\textwidth]{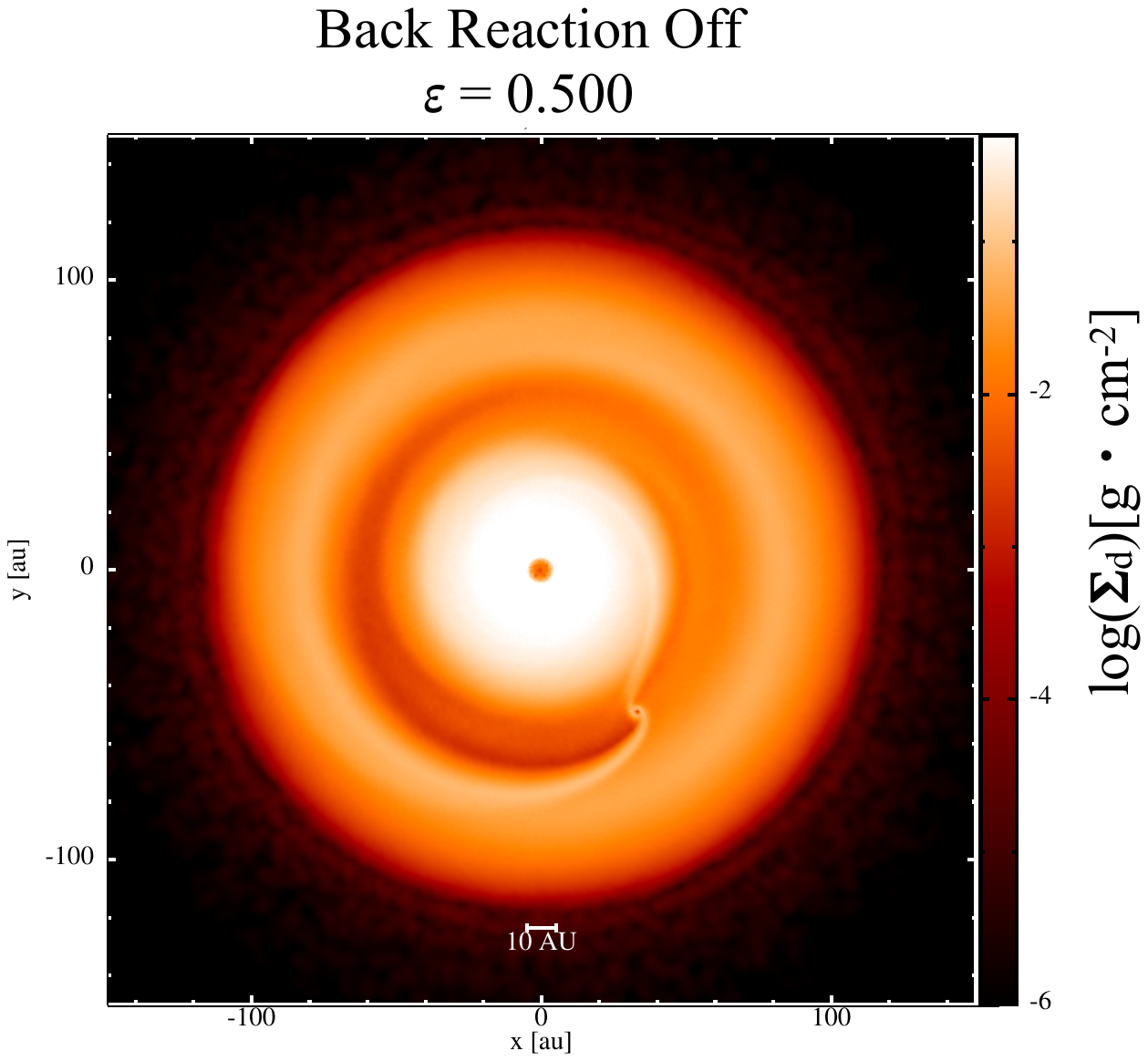} \\
    \end{tabular}
   \caption{Dust surface density of simulated discs with $a=0.1$mm. The left column is simulated with the back reaction, and the right column is simulated without the back reaction. The top row is $\epsilon = 0.10$ and the bottom row is $\epsilon = 0.50$. The disc that includes the back reaction and with $\epsilon=0.50$ has the least prominent gap of any of the discs simulated and also has smoothest overall distribution. Additionally, in these high $\epsilon$ simulation, the role of $\epsilon$ and the back reaction in gap closing is readily apparent. Simulations with the back reaction have gaps spanning a smaller azimuthal angle over all}
    \label{fig:Sarracen2}
\end{figure*}

\section{Results}\label{Results}
\begin{figure*}
    \centering
    \includegraphics[width=\textwidth]{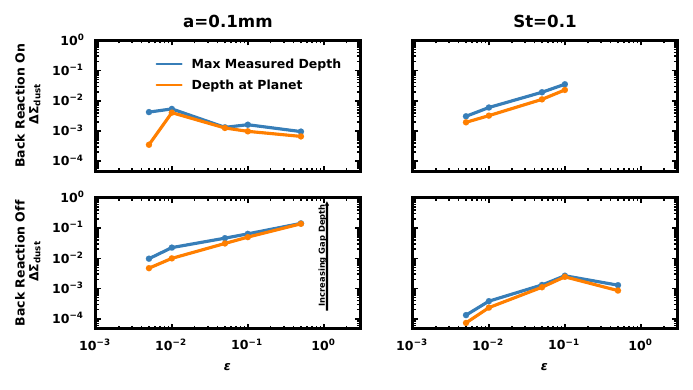}
    \caption{Measured gap depth at the planet location and gap maximum for all simulations. The top row is $\Delta\Sigma_\mathrm{dust}$ with the back reaction on and the bottom row is $\Delta\Sigma_\mathrm{dust}$ with the back reaction off. The left column shows results for constant grain size of a = $0.1$mm, and the right column shows results for constant St number of $0.1$. Data sets for St=0.1 with the back reaction is missing a point at $\epsilon=0.50$.}
    \label{fig:QuadPlot}
\end{figure*}

Figure \ref{fig:Sarracen1} and  Figure \ref{fig:Sarracen2} show the surface density plots of the constant grains simulations, with back reaction included in the left hand column, and turned off in the right hand column.  $\epsilon$ increases from top to bottom and from Figure \ref{fig:Sarracen1} to Figure \ref{fig:Sarracen2}. It is clear that simulating the back reaction substantially alters the observed dust morphology - in particular, the degree of 2D axisymmetry displayed. We therefore suggest that in modelling the dusty components of discs (such as for synthetic ALMA observations), that the dust back reaction is always included.  

In Figure \ref{fig:Sarracen1} to Figure \ref{fig:Sarracen2} we also see the back reaction enhancing pressure trapping of grains around the co-orbital region of the planet at low $\epsilon$. At high $\epsilon$ the back reaction contributes to gap infilling, potentially hindering gap clearing in the first place. Without the back reaction, relative surface densities of the discs indicate that pressure trapping around the gap is minimised as dust seems to readily move towards the central star; in the inner $5-10\mathrm{AU}$ surface densities begin to approach unity. Still there is a prominent gap in the co-orbital region of the planet, and it deepens steadily compared to the background behaviour of the disc as $\epsilon$ increases. 

Surface density plots for all constant St simulations are presented in Appendix \ref{StokesFigs}. 

\subsection{Depth vs $\epsilon$}

\subsubsection{Constant Grain Size}\label{DepthGrain}

Our primary results are shown in Figure \ref{fig:QuadPlot}, 
which demonstrates that there is a clear relationship between gap depth and $\epsilon$ in all scenarios, suggesting that apparent dust morphology in ALMA observations could potentially be used as a tool to constrain $\epsilon$.

In the $a = 0.1$ mm simulations, when the back reaction is included, increasing $\epsilon$ results in an increased gap depth from $\epsilon=0.005$ to $\epsilon=0.01$. As $\epsilon$ increases beyond this, the gap depth decreases. This might suggest that there is a sweet spot for $\epsilon$ where maximum gap depth can be achieved. This also seems to indicate a consistent regime from $\epsilon = 0.05$ - $0.50$, but that space is very coarsely sampled.

When the back reaction is not included, gap depth increases monotonically with $\epsilon$ for the $a=0.1$ mm scenario. This result is consistent with previous findings (e.g. \citealt{dipierro_gas_2018}). This demonstrates that neglecting the back reaction when using models to interpret observations is likely to result in incorrect conclusions. Gap depth continues to increase with increasing $\epsilon$.  

\subsubsection{Constant Stokes Number}\label{DepthStokes}

We simulate discs of a constant St number of $0.1$ with all other parameters described in Section \ref{SPHModel}. The right column of Figure \ref{fig:QuadPlot} presents the two different measures of depth versus $\epsilon$, with and without the back reaction. 

With the back reaction included, we see a trend of gap depth increasing as $\epsilon$ increases (top right of Figure \ref{fig:QuadPlot}). When the back reaction is not included, we see a similar trend of gap depths increasing until $\epsilon = 0.10$, where there is a decrease in gap depths at $\epsilon = 0.50$ (bottom right Figure \ref{fig:QuadPlot}). This indicates a potential regime change from "increasing $\epsilon$ increases gap depth" to "increasing $\epsilon$ decreases gap depth" in the range $\epsilon = 0.10 - 0.50$. Due to $St=0.1$ with $\epsilon=0.50$ being computationally prohibitive when the back reaction is included, we do not know if this is due to the lack of the back reaction specifically or changing $\epsilon$ in general.  

\subsection{Reduced Profiles}
\begin{figure*}
    
    \centering
    \begin{tabular}{cc}
     \includegraphics[width=0.5\textwidth]{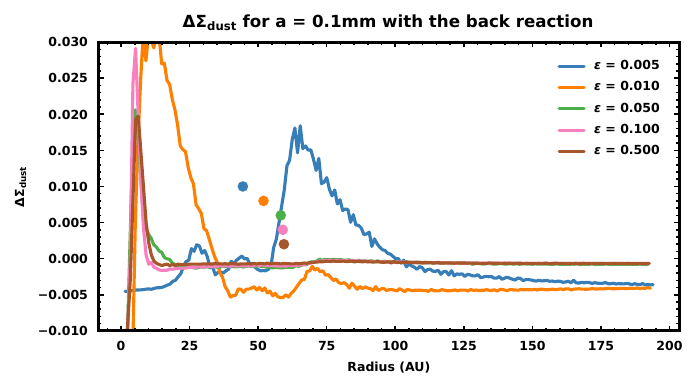} \includegraphics[width=0.5\textwidth]{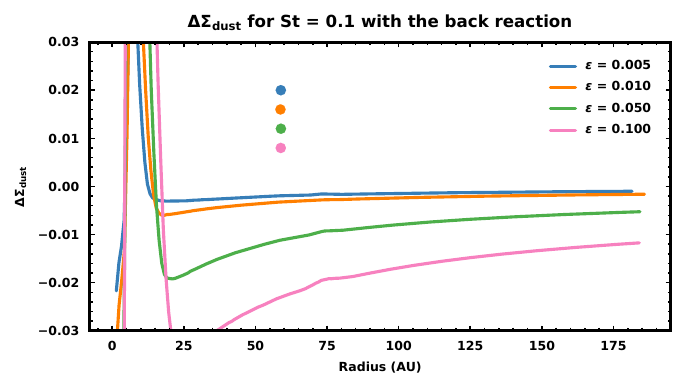}  \\
      \includegraphics[width=0.5\textwidth]{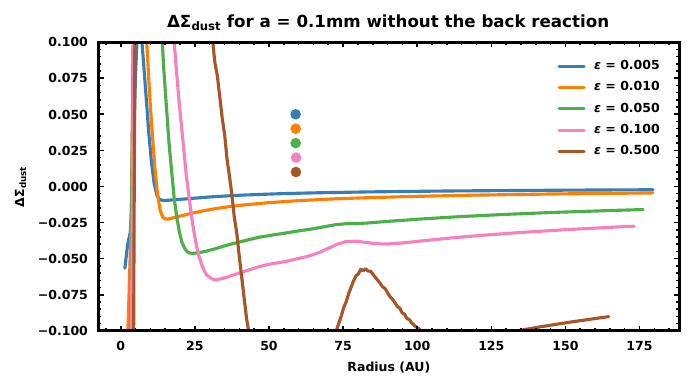}   \includegraphics[width=0.5\textwidth]{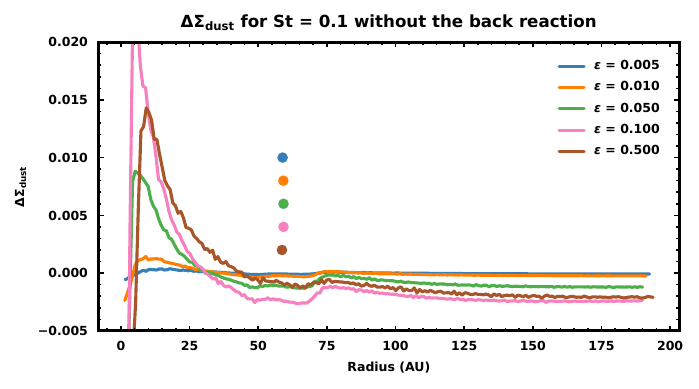}
    
    \end{tabular}
    \caption{Reduced surface densities of simulations with constant grain size of $a=0.1$ mm (left column) and constant St number of $0.1$ (right column). The location of each planet is plotted at its radius in the same color as the associated profile. Vertical displacement is solely to prevent overplotting in the case of planets being at approximately the same radius. In the bottom left plot, some behaviour of the $\epsilon=0.50$ profile is omitted to better visual more shallow gap profiles.}
    \label{fig:ReducedDen}
\end{figure*}
\subsubsection{Constant Grain Size}\label{GrainProf}

We present the reduced surface density profiles used for the "Depth vs. $\epsilon$" plots in Section \ref{DepthGrain} (top left of Figure \ref{fig:ReducedDen}). Notable here is the exterior/interior pressure trapping \citep{dipierro_dust_2015} of dust at $\epsilon = 0.005$ and $\epsilon = 0.010$ respectively. Additionally the depth regime transition after $\epsilon = 0.010$ is easily identified. At low $\epsilon$ we also see the soft "w" shape initially described in Section \ref{Gaps}, indicating there is some dust co-orbiting with the planet. This shape appears in the profile as pronounced inner and outer edges relative to the gap depth, and a small increase of surface density co-located with the planet that appears as a small bump. The lowest $\epsilon$ has the most co-orbital dust and exterior pressure trapping, suggesting the back reaction is insufficient in overcoming radial drift barriers. 
In the case with the back reaction, \begin{referee_comments}
    we see planetary migration on the order of $15 \mathrm{AU}$ at the same low $\epsilon$ that produces the soft "w" shape. This amount of planetary migration over 80 orbits is consistent with migration rates and timescales seen in previous SPH investigations of protoplanetary discs \citep{schafer_simulations_2004, benitez-llambay_long-term_2016, rowther_planet_2020}. What is notable is that we only see migration in our simulations when the back reaction is accounted for, and at low $\epsilon$. The behaviour at low $\epsilon$ is likely due to our keeping dust mass constant. As $\epsilon$ increases, gas mass must necessarily decrease, and reaches a point where mass interior to the planet is insufficient in driving migration \citep{quillen_planet_2004,crida_cavity_2007}.
\end{referee_comments}

In the high $\epsilon$ regime when the back reaction is present, gaps almost entirely lose that soft "w" shape, being able to be differentiated only by slight changes in depth. This behaviour is different from the bottom left of Figure \ref{fig:ReducedDen}, constant grain size without the back reaction, where surface density in the gap decreases dramatically with increasing $\epsilon$. This is an indication of the role of the back reaction in gap formation .   

\subsubsection{Constant Stokes Number}\label{StokesProf}
For a constant St number of $0.1$, motion is dominated by the gas phase with dust responding according to drag forces. By direct comparison between the reduced surface densities with and without the back reaction, we notice a "smoothing" effect of the back reaction that seems to be accentuated when drag forces are kept constant. Compared to simulations without the back reaction, the profiles in the top right plot of Figure \ref{fig:ReducedDen} lack small scale variation in the surface density with respect to $R$, being qualitatively smooth. In the bottom right of Figure \ref{fig:ReducedDen} we see that the simulated profiles have much more small scale variation in the surface density. Additionally, in simulations without the back reaction, there is a soft, sloping "w" at high $\epsilon$, most clearly observable in $\epsilon= 0.10$ but also marginally visible in $\epsilon = 0.50$. 

When considering gap depth, it is interesting to note that trends with and without the back reaction are inverse for the two scenarios. For constant grain sizes, gap depth increases monotonically with $\epsilon$ when the back reaction is absent. The inverse is true for constant St number: when the back reaction is present, gap depths increase monotonically with $\epsilon$ . 

\subsection{The Effect of the Back Reaction on Gap Depth}\label{BackDepth}

\begin{figure*}
    \centering
    \includegraphics[width=\textwidth]{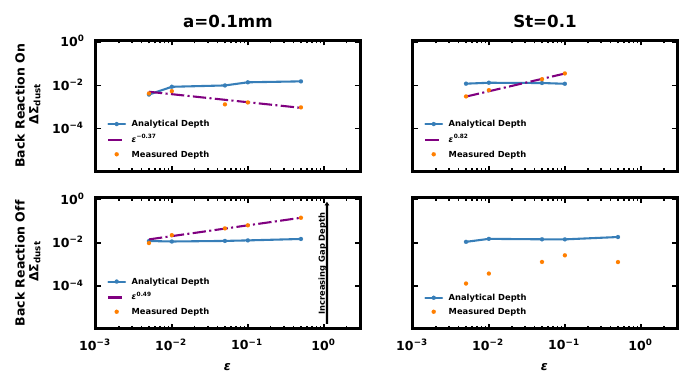}
    \caption{Measured gap depth in dust density compared to the analytical prescription from \protect\cite{tanaka_eccentric_2022} (Equation \eqref{eq:estimatedGap}). In the simulations with constant grain size, gaps in dust are suppressed by the back reaction as $\epsilon$ increases (top left). Without the back reaction, gaps increase as $\epsilon$ increases (bottom left). Neither has good agreement with the analytical prescription. In the simulations with constant St number, there is an apparent regime transition around 
    $\epsilon=0.10$ without the back reaction (bottom right).
    Unfortunately, $\epsilon=0.50$ was computationally prohibitive with the back reaction on, so it is unclear if that regime transition would be present there as well (top right).
    Overplotted in all plots except the bottom right is a fit of $\epsilon^{\xi}$ to the measured data. The value of $\xi$ is presented in the legend of the related scenario and characterizes how measured gap depths change with respect $\epsilon$ for a given scenario. 
    }
    \label{fig:AnalyticalDust}
\end{figure*}

We demonstrate that including the back reaction distinctly alters the disc morphology from extant prescriptions (Figure \ref{fig:AnalyticalDust}). For the constant grain size scenario (top left of Figure \ref{fig:AnalyticalDust}), including the back reaction results in dust gap depth decreasing with $\epsilon$. If we turn off the back reaction (bottom left of Figure \ref{fig:AnalyticalDust}), the dust gap depth increases with $\epsilon$. \begin{referee_comments}
    We suspect this this is due to the back reaction increasing pressure trapping at the edges of the gap. When including the back reaction, $\Lambda_\mathrm{net}$ felt by dust decreases with $\epsilon$, and this effect is especially pronounced in regions of the disc with high local $\epsilon$. This yields the inner and outer disc edges that appear as a soft ''w'' shape at globally low $\epsilon$. The $n^\mathrm{th}$ parcel of dust swept away from the planet via deposition torque experiences marginally less $\Lambda_\mathrm{net}$ compared to the $n-1^\mathrm{th}$ parcel of dust due to a slightly higher local $\epsilon$, and begins to pile-up at the gap edge. This process repeats for the $n+1^\mathrm{th}$ parcel of dust experiencing a still smaller $\Lambda_\mathrm{net}$ Without the back reaction, all parcels of dust feel the same $\Lambda_\mathrm{net}$ and thus are cleared much more uniformly.
\end{referee_comments}

A slightly different discrepancy is seen in the constant $\mathrm{St}=0.1$ scenario. When the back reaction is simulated (top right of Figure \ref{fig:AnalyticalDust}), measured gap depths are more shallow than the analytical prescription, starting at $\Delta\Sigma_\mathrm{dust}\approx2\cdot10^{-3}$, before rising above the analytical prescription to $\Delta\Sigma_\mathrm{dust}\approx4\cdot10^{-2}$. The analytical prescription maintains a roughly constant $\Delta\Sigma_\mathrm{dust}\approx1\cdot10^{-2}$. When the back reaction is off (bottom right of Figure \ref{fig:AnalyticalDust}), the analytical prescription remains roughly constant on the order of $10^{-2}$ while measured depths are one to two orders of magnitude below that for all values of $\epsilon$. This demonstrates that inclusion of the back reaction, even when simulating PPDs with small values of $\epsilon$, may be necessary to fully understand planet-disc interactions (see for example \citet{guilera_quantifying_2023}). 
Additionally, explicit consideration of the back reaction can be used to constrain $\epsilon$ from observations of gap morphology. As it stands our analysis of depths can be used to create a system of relative ranking of $\epsilon$ by examining gap depth relative to an unperturbed disc. Unfortunately analysis of gap depths depends on many factors including planetary mass and $\epsilon$, so discs with planets of well known or well constrained mass are needed for this ranking to be effective. That being said, for a number of planets whose masses are similar and constrained through kinematics, one could rank the $\epsilon_i$ of each of the host discs through observations of the gaps induced by planetary action, e.g.: $\epsilon_{B} \leq \epsilon_{C} << \epsilon_{A}$, for some selection of discs found to be hosting planets of comparable mass. In conjunction with continuum observations, this also makes disc mass directly comparable. 

With these simulations, we have demonstrated that for constant planetary mass, gap depth is dependent on $\epsilon$ and the back reaction. If a few observed discs have masses that are well constrained from other means, it may be possible to straightforwardly derive a "mass ladder" from direct comparison of dust continuum flux. 

 \subsection{Comparison to an Analytical Depth Criterion}\label{AnalyticalDepth}
\begin{figure*}
    \centering
    \includegraphics[width=\textwidth]{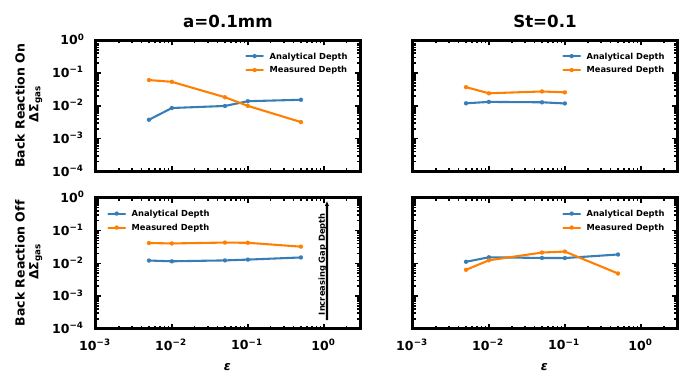}
    \caption{Measured gap depth in gas density compared to the analytical criterion from \protect\cite{tanaka_eccentric_2022} (Equation \eqref{eq:estimatedGap}). In the simulations with constant grain size, gaps in gas are suppressed by the back reaction as $\epsilon$ increases (top left). In all other scenarios, the gap depth is mostly constant. The only scenario to see numerical agreement with the analytical criterion are the simulations with constant St number and without the back reaction. This scenario is, to a good approximation, the regime under which the criterion was developed: well coupled dust that largely follows the gas, without the back reaction.}
    \label{fig:AnalyticalGas}
\end{figure*}

We solved for an analytical gap depth estimation provided by Equation \eqref{eq:estimatedGap} for each disc. We then made a direct comparison to the gap depth measurement scheme discussed in Section \ref{Gaps} by shifting Equation \eqref{eq:estimatedGap} into $\Delta\Sigma$ space. This is accomplished by taking
\begin{equation}
    \Delta \Sigma = 1-\frac{\Sigma_\mathrm{p}}{\Sigma_0}, 
\end{equation}
and assumes that for a surface density profile with no gap, $\Sigma_\mathrm{No Gap} = \Sigma_0$. We perform this calculation, and compare it to our measures for $\Delta\Sigma_\mathrm{dust}$ and $\Delta\Sigma_\mathrm{gas}$ in all scenarios, with and without the back reaction. Results are presented in Figure \ref{fig:AnalyticalDust} for the dust phase, and Figure \ref{fig:AnalyticalGas} for the gas phase. The scale heights of each simulation, $h_\mathrm{p}$, measured at the planetary radius are all relatively constant at $\sim2.6$ au, not changing directly as a strong function of $\epsilon$. \begin{referee_comments}
Of note, Equation \eqref{eq:estimatedGap} was developed for application to the gas phase, but in scenarios where dust is well-coupled and gas does not experience the back reaction \citep{tanaka_eccentric_2022}. We compare the gas gap depths predicted by Equation \eqref{eq:estimatedGap} to $\Delta\Sigma_\mathrm{dust}$ to show agreement or contradiction with the overall trends in gap depth as a function of $\epsilon$.
\end{referee_comments}

In a contradiction with Equation \eqref{eq:estimatedGap} and \citet{tanaka_eccentric_2022}, we find that the depth of the gas gap affected by changes in $\epsilon$. We also see a noticeable dependence on $\epsilon$ for measured dust gap depths. For each set of simulations, we examine the dust phase, and fit a power law function, $\epsilon^{\xi}$ where $\xi$ is a fitting constant, unique and fitted to each scenario.
\begin{align}
\Delta\Sigma_\mathrm{dust} &\propto \epsilon^{\xi}.
\end{align}
We present a plot of measured gap depths, analytical gap depths, and $\epsilon^{\xi}$ in Figure \ref{fig:AnalyticalDust}. All scenarios are distinct from each other as functions of $\epsilon$, and no simulation set has the same behaviour as the analytical prescription.

For the scenario of constant grain size and back reaction, initially considering only the dust, the analytical estimation from Equation \eqref{eq:estimatedGap} and our measured gap depths start with some agreement at $\epsilon=0.005$ and $\epsilon=0.01$, but diverge as $\epsilon$ increases to $0.10$ and beyond. A possible explanation are the length scales of $K'$ changing. The inward migration seen by planets for low values of $\epsilon$ may be meaningfully changing the analytically determined gap depth via the factor of $r_\mathrm{p}^{5}$ in $K'$ (Equation \eqref{eq:KPrime}). Since $h_\mathrm{p}$ as a function of radius is roughly constant between simulations, the planetary migration that is only seen for $\epsilon=0.005$ and $\epsilon=0.01$ results in length scales that bring our measured gap depths into agreement with the analytical gap depth. 
\begin{referee_comments}
    The planetary migration is happening on time and length scales that are consistent with previous work \citep{benitez-llambay_long-term_2016, schafer_simulations_2004, rowther_planet_2020},
\end{referee_comments} but it is unclear if bringing the analytical and measured depths into agreement is incidental or related to the mechanisms that are driving migration in the first place. For this reason, we do not put much significance in the trend with respect to $\epsilon$ in these simulations. 

In the simulations with constant St without dust back reaction, gap depth of the dust increases with increasing $\epsilon$, but the largest value of epsilon reverses this trend. 
Due to $\epsilon = 0.50$ being computationally prohibitive to simulate for constant St with the back reaction, we cannot say if this trend is something that is or is not suppressed by the back reaction. 

In a dusty, viscous, pressure-supported protoplanetary disc, torque excited in the gas by the presence of a planet will be communicated to the dust via drag interactions, with the drag torque from gas onto dust given by Equation \eqref{eq:drag1}. At the same time, dust exerts a torque back on the gas via mutual drag (Equation \eqref{eq:BackRe1}). This is back reaction and is proportional to the dust-to-mass ratio, $\epsilon$. Comparison to the analytical depth criterion of \citet{tanaka_eccentric_2022} reveals that there is a dependence on $\epsilon$ between $\epsilon^{0.5}$ and $\epsilon^{1.0}$ that is lacking from current theory. 

While not as straightforward as the ranking method described in Section \ref{BackDepth}, a related analytical method allows for more precise estimates of $\epsilon$ from observations. Expanding the work of \citet{kanagawa_formation_2015} and \citet{tanaka_eccentric_2022} into a regime that considers the back reaction due to dust provides another analytical tool to constrain $\epsilon$ and PPD masses from observations.

\section{Conclusions and Future Work}\label{Conclusion}

\subsection{Conclusions}

We ran a series of SPH simulations with and without the back reaction of dust and varying $\epsilon$ to investigate the effects of dust on disc morphology. We measure gap depths by comparing each simulated scenario to a fitted logarithmic surface density taken to be the relaxed scenario without a planet. We find that the effect of the back reaction on gap formation due to planetary action is significant in both gas and dust phases, and conventional prescriptions of gap formation need updating to account for the effects of the back reaction. 

\begin{referee_comments}
    When grain size is kept constant and the back reaction is included, gap depths induced in the dust phase tend to decrease as $\epsilon$ increases. This is contrary to what is found in scenarios of fixed grain size where the back reaction is not included (Left column in Figure \ref{fig:AnalyticalDust}). We propose that not including the back reaction in simulations leads to incorrect results. 
    Additionally, an observational prediction of this result is that as $\epsilon$ increases in observed discs, we expect the measured brightness drop between the gap region and the rest of the disc to decrease in intensity. 
    When the Stokes number of the dust is kept constant, we see that gap depths in the dust phase increase as $\epsilon$ increases, with and without the back reaction. This result does not map directly to observational predictions, since a constant Stokes number yields varying grain sizes, and therefore emission intensity, at a given location in the disc. 
    We also compare the gap depths formed in gas to the analytical prescription outlined by \citet{tanaka_eccentric_2022} (Eq. \ref{eq:estimatedGap} and Figure \ref{fig:AnalyticalGas}), using the same gap depth measuring scheme described for the dust phase above. Our findings suggest the overall importance of the back reaction for the gas phase in addition to the dust. When including the back reaction and fixing grain size (top left of Figure \ref{fig:AnalyticalGas}), gap depths measured in gas follow the same trend as gap depths measured in dust: decreasing in depth as $\epsilon$ increases. 
    In the scenario where the back reaction is not included, gaps induced in gas appear to have no or a very slight trend with $\epsilon$.
    
    We conclude that considering the back reaction when simulating a dusty, viscous, pressure supported protoplanetary disc is crucial, as the effect on planet induced gap formation is significant in both the dust and gas phases.  
\end{referee_comments}

\subsection{Future Work}
We have demonstrated a relationship between $\epsilon$ and gap depths that is not fully accommodated by Equation \eqref{eq:estimatedGap} and the lineage of \citet{tanaka_eccentric_2022}, but more work is required to refine an analytical relationship between $\epsilon$ and the gap depth. Our $\epsilon$ parameter space was coarsely sampled to demonstrate the overall impact and importance of the effect, but we lack sufficient data to generate a robust model. We suggest that future analysis of the back reaction focus in the regime of $\epsilon \in [0.005,0.01,0.05]$ as that is the interval over which the $a=0.01$ mm scenario transitions from increasing gap depth to decreasing gap depth.

Additionally, holding $\epsilon$ constant and varying planetary mass could demonstrate the effect of the back reaction on gap formation through direct comparison to depths estimated from Equation \eqref{eq:estimatedGap}. This also could be used to provide either a corrective term or an explicit $\epsilon$ dependence to the analytical depth formula. 
In the presence of planets, we have demonstrated that considering the effects of dust and the back reaction introduces significant variation when $\epsilon$ is poorly constrained. Gravitational instability, magnetic effects, and/or streaming instability may similarly respond to changing $\epsilon$ and the back reaction in a way that makes constraint or identification more straightforward. 

A secondary or tertiary effect beyond the scope of this study is the planetary migration seen in the scenario of constant grain size with the back reaction on. The simulations in which we see the most planetary migration are those with the most mass in gas, and as $\epsilon$ increases, total gas mass must necessarily go down to keep dust mass constant. This is consistent with exploration into planetary migration by \cite{quillen_planet_2004} and \cite{crida_cavity_2007} where it was discussed that planets could not migrate inwards if the planet itself was more massive than the interior gas mass. 

A recent study from \cite{guilera_quantifying_2023} provides some additional context to planetary migration under changing $\epsilon$ by explicitly considering torque contributions from dust components. They are able to change the sign and magnitude of net torques in the regime of low-mass planets ($M_\mathrm{p} \leq 10M_\oplus$) when considering different mass ratios and coupling regimes. We suspect that findings from \cite{guilera_quantifying_2023} are related to the variation we see in migration, slowing as $\epsilon$ increases. Despite our planetary masses not being in the same regime, there is a congruent finding: an increase of $\epsilon$ leading to slowed migration. We suspect this is due to an increase in positive torques from dust. Extending the work of \cite{guilera_quantifying_2023} to $2M_\mathrm{Jup}$ planets will make this hypothesis testable. We suggest a grid-based code; SPH as an algorithm has difficulty with the long timescales typically associated with planetary migration.

\section*{Acknowledgements}

Special thanks to Jess Speedie for insights into observational morphology that resulted in measuring of gap depth relative to the background surface density profile. 

Data visualisation performed in Figures \ref{fig:Sarracen1}, \ref{fig:Sarracen2}, and Appendix \ref{StokesFigs} was made possible by SPLASH, an SPH visualisation utility \citep{price_splash_2007}. Dr. Cassandra Hall acknowledges support from NSF AAG grant No. 2407679, the Georgia Museum of Natural History, and the National Geographic Society.

This study was supported in part by resources and technical expertise from the Georgia Advanced Computing Resource Center, a partnership between the University of Georgia’s Office of the Vice President for Research and Office of the Vice President for Information Technology. 
\section*{Data Availability}\label{DataAvail}
Data used in this study and all files necessary to reproduce the simulations discussed can be acquired upon reasonable request by contacting Dr. Cassandra Hall at the University of Georgia's Center for Simulational Physics.


\bibliographystyle{mnras}
\bibliography{example}



\appendix
\section{Fitting Parameters for Constant Grain Size and Constant Stokes Number}\label{zetaGas}
A report of the fitting parameter, $\zeta$, from Equation \ref{eq:ZETA} for each of the discs simulated with constant grain size and constant Stokes number. See Tables \ref{tab:zetaGrainY} and \ref{tab:zetaGrainN} for the simulations of constant grain size with and without the back reaction, respectively. See Tables \ref{tab:zetaStokesY} and \ref{tab:zetaStokesN} for the simulations of constant Stokes number with and without the back reaction, respectively. 

\begin{table}
    \centering
    \caption{Table of fitted $\zeta$ following Equation \ref{eq:ZETA} for simulations of constant grain size that included the back reaction.}
    \begin{tabular}{l c|c|c}
        \hline
         $\epsilon$ & $\zeta_\mathrm{dust}$ & $\zeta_\mathrm{gas}$ \\
         \hline \hline
         $0.005$ & -0.02 & -0.48\\
         $0.01$  & -0.48 & -0.47\\
         $0.05$  & -0.59 & -0.48\\
         $0.1$   & -0.64 & -0.48\\
         $0.50$  & -0.60 & -0.45\\
        \hline
    \end{tabular}
    \label{tab:zetaGrainY}
\end{table}

\begin{table}
    \centering
    \caption{Table of fitted $\zeta$ following Equation \ref{eq:ZETA} for simulations of constant grain size that did NOT include the back reaction.}
    \begin{tabular}{l c|c|c}
        \hline
         $\epsilon$ & $\zeta_\mathrm{dust}$ & $\zeta_\mathrm{gas}$  \\
         \hline \hline
         $0.005$ & -0.66 & -0.44\\
         $0.01$  & -0.72 & -0.45\\
         $0.05$  & -0.62 & -0.41\\
         $0.1$   & -0.59 & -0.40\\
         $0.50$  & -0.50 & -0.39\\
        \hline
    \end{tabular}
    \label{tab:zetaGrainN}
\end{table}

\begin{table}
    \centering
    \caption{Table of fitted $\zeta$ following Equation \ref{eq:ZETA} for simulations of constant Stokes number that included the back reaction. The simulation for $\epsilon$=0.50 was not completed and therefore not included.}
    \begin{tabular}{l c|c|c}
        \hline
         $\epsilon$ & $\zeta_\mathrm{dust}$ & $\zeta_\mathrm{gas}$  \\
         \hline \hline
         $0.005$ & -0.64 & -0.45\\
         $0.01$  & -0.63 & -0.43\\
         $0.05$  & -0.68 & -0/45\\
         $0.1$   & -0.60 & -0.41\\
         $0.50$  & N/A & N/A\\
        \hline
    \end{tabular}
    \label{tab:zetaStokesY}
\end{table}

\begin{table}
    \centering
    \caption{Table of fitted $\zeta$ following Equation \ref{eq:ZETA} for simulations of constant Stokes number that did NOT include the back reaction}
    \begin{tabular}{l c|c|c}
        \hline
         $\epsilon$ & $\zeta_\mathrm{dust}$ & $\zeta_\mathrm{gas}$  \\
         \hline \hline
         $0.005$ & -0.35 & -0.48\\
         $0.01$  & -0.43 & -0.50\\
         $0.05$  & -0.52 & -0.54\\
         $0.1$   & -0.55 & -0.54\\
         $0.50$  & -0.44 & -0.46\\
        \hline
    \end{tabular}
    \label{tab:zetaStokesN}
\end{table}

\section{Stokes with and without the back reaction}\label{StokesFigs}
\begin{figure*}
    \centering
    \begin{tabular}{cc}
        \includegraphics[width=0.5\textwidth]{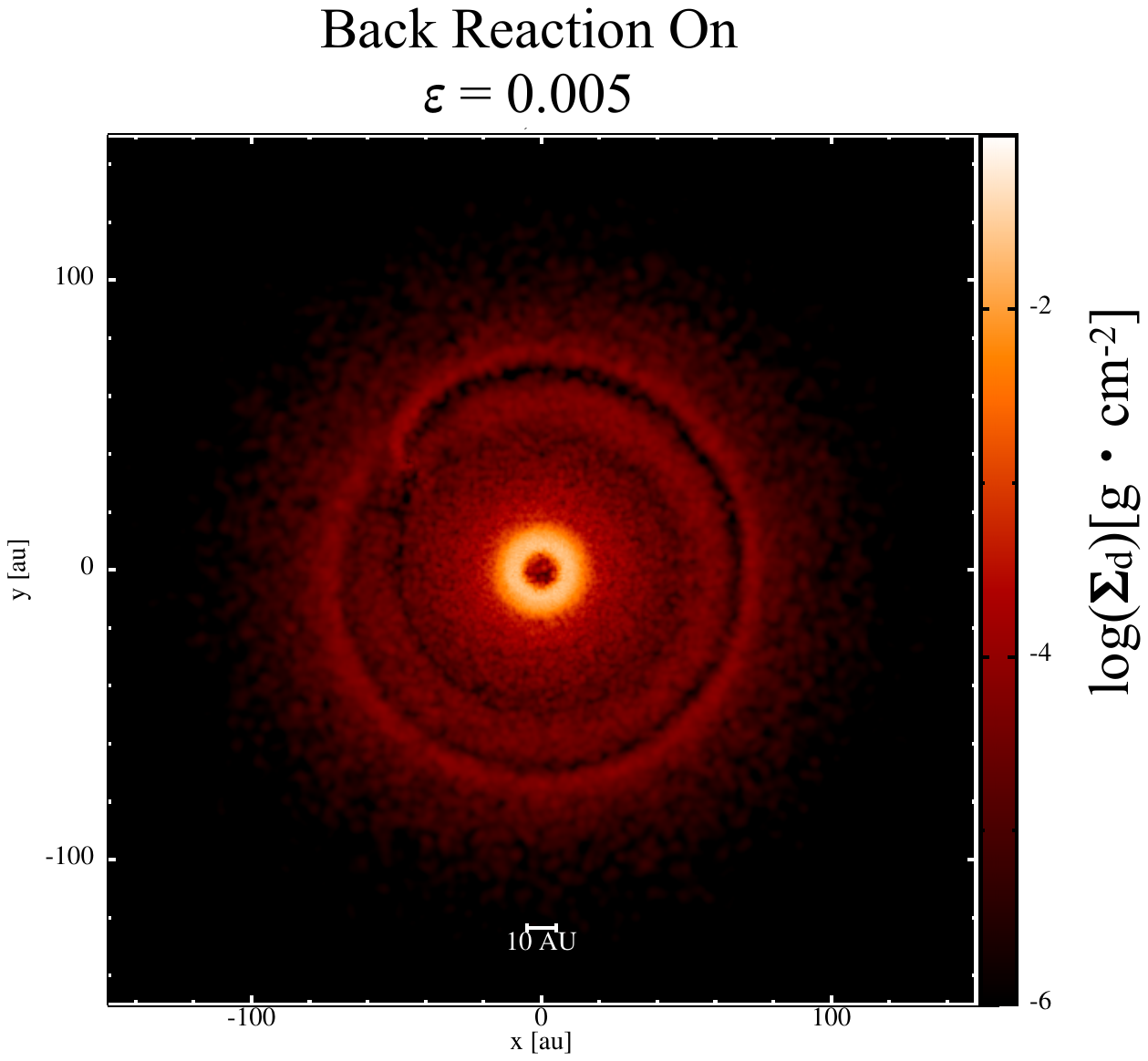} & \includegraphics[width=0.5\textwidth]{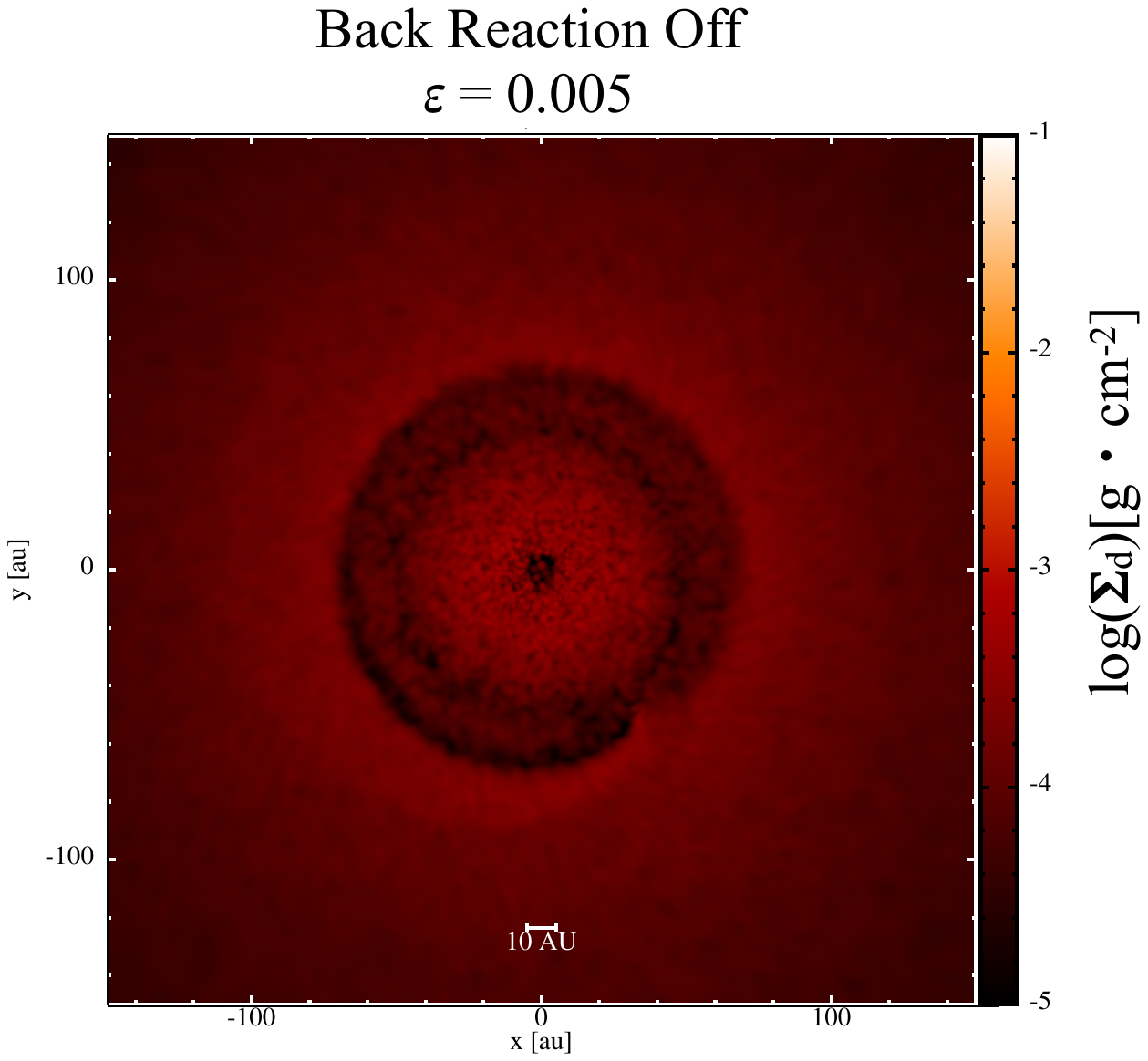} \\
        \includegraphics[width=0.5\textwidth]{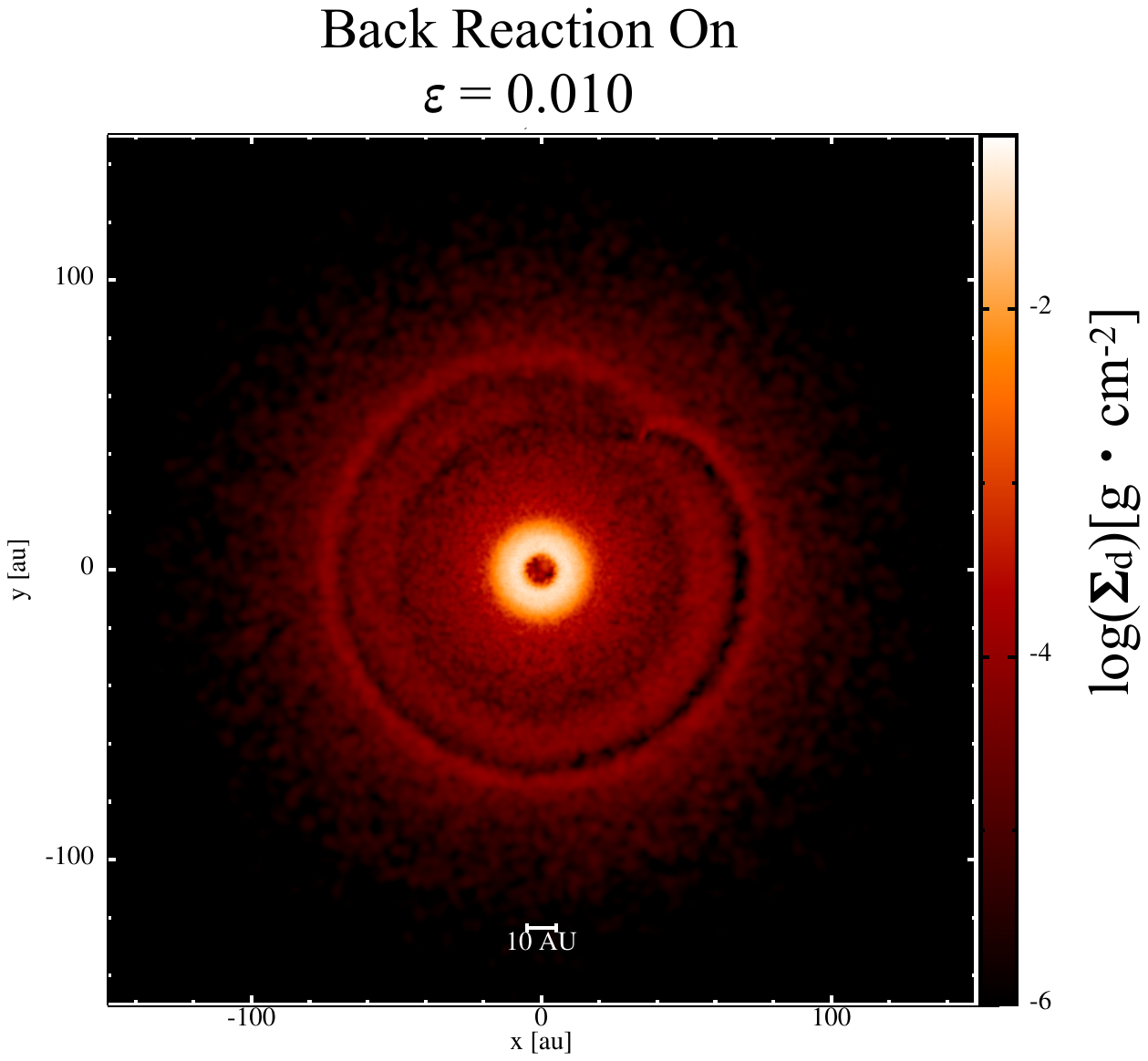} & \includegraphics[width=0.5\textwidth]{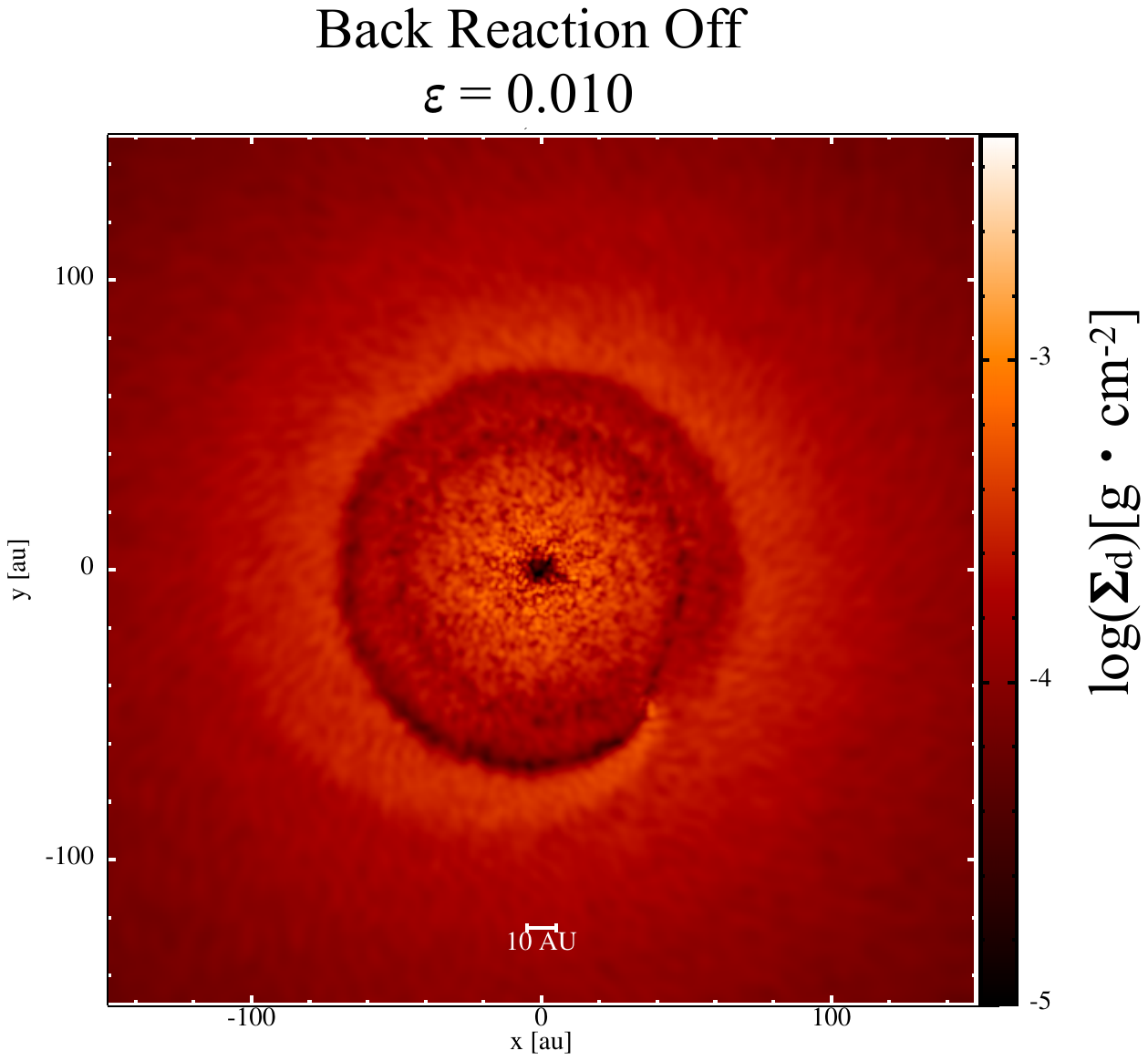} \\
        \includegraphics[width=0.5\textwidth]{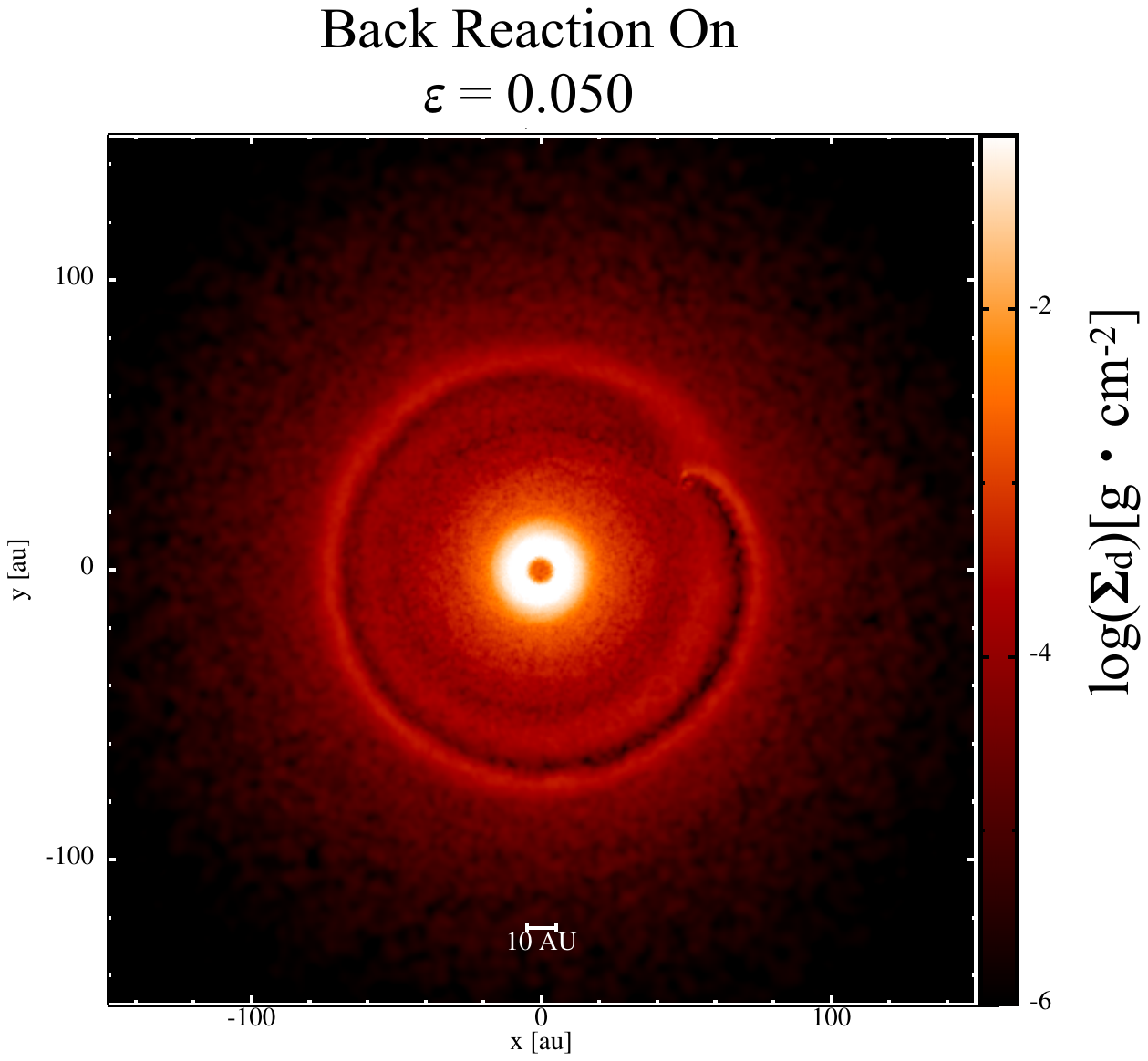} & \includegraphics[width=0.5\textwidth]{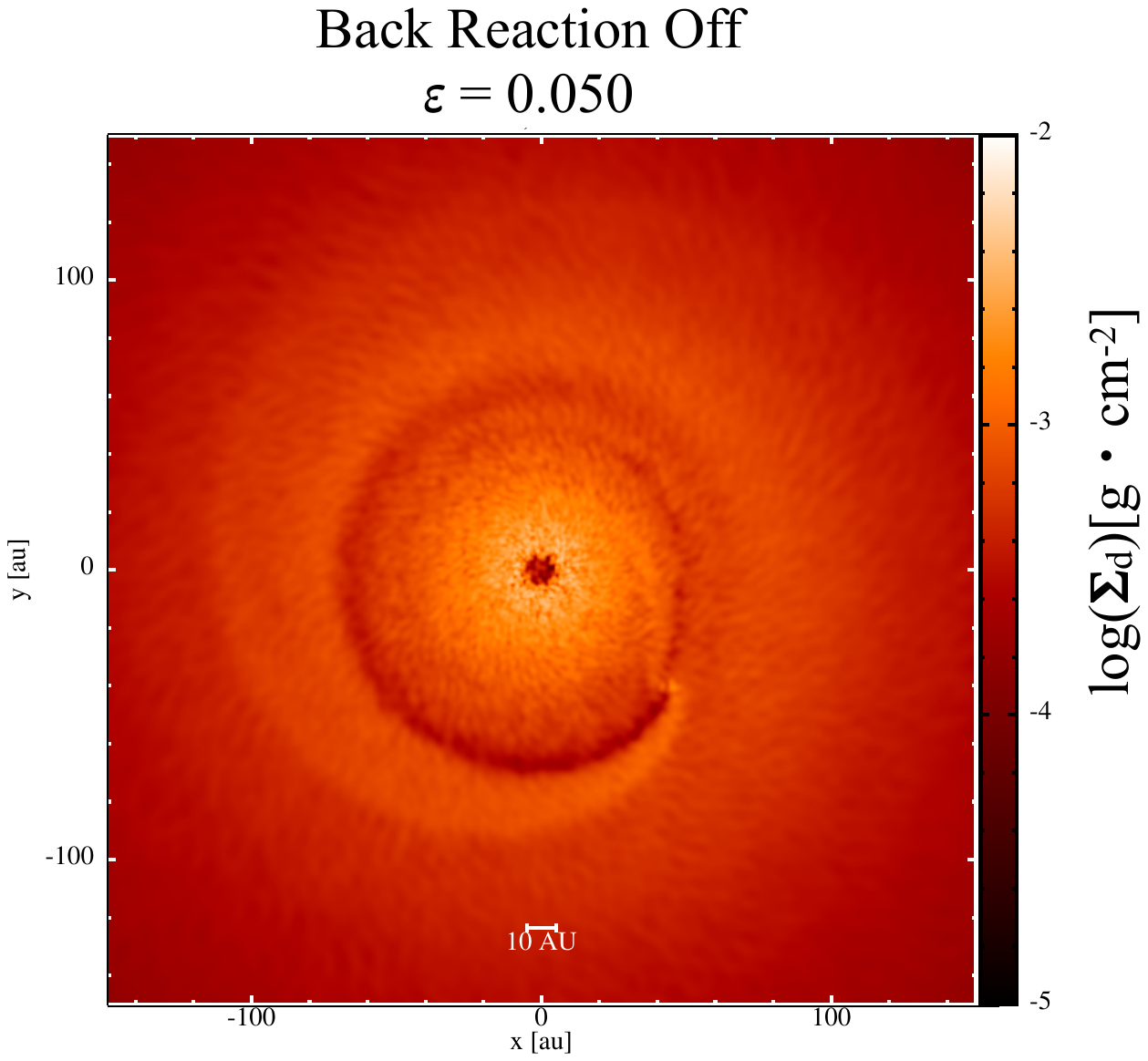} \\
    \end{tabular}
     \caption{Dust surface density of simulated discs with a constant Stokes Number of $\mathrm{St}=0.1$. The left column is simulated with the back reaction, and the right column is simulated without the back reaction. Each row is a different $\epsilon$, increasing down the page and into Figure \ref{fig:StokesDen2}.}
     \label{fig:StokesDen1}
    \end{figure*}
    \begin{figure*}
    \begin{tabular}{cc}
        \includegraphics[width=0.5\textwidth]{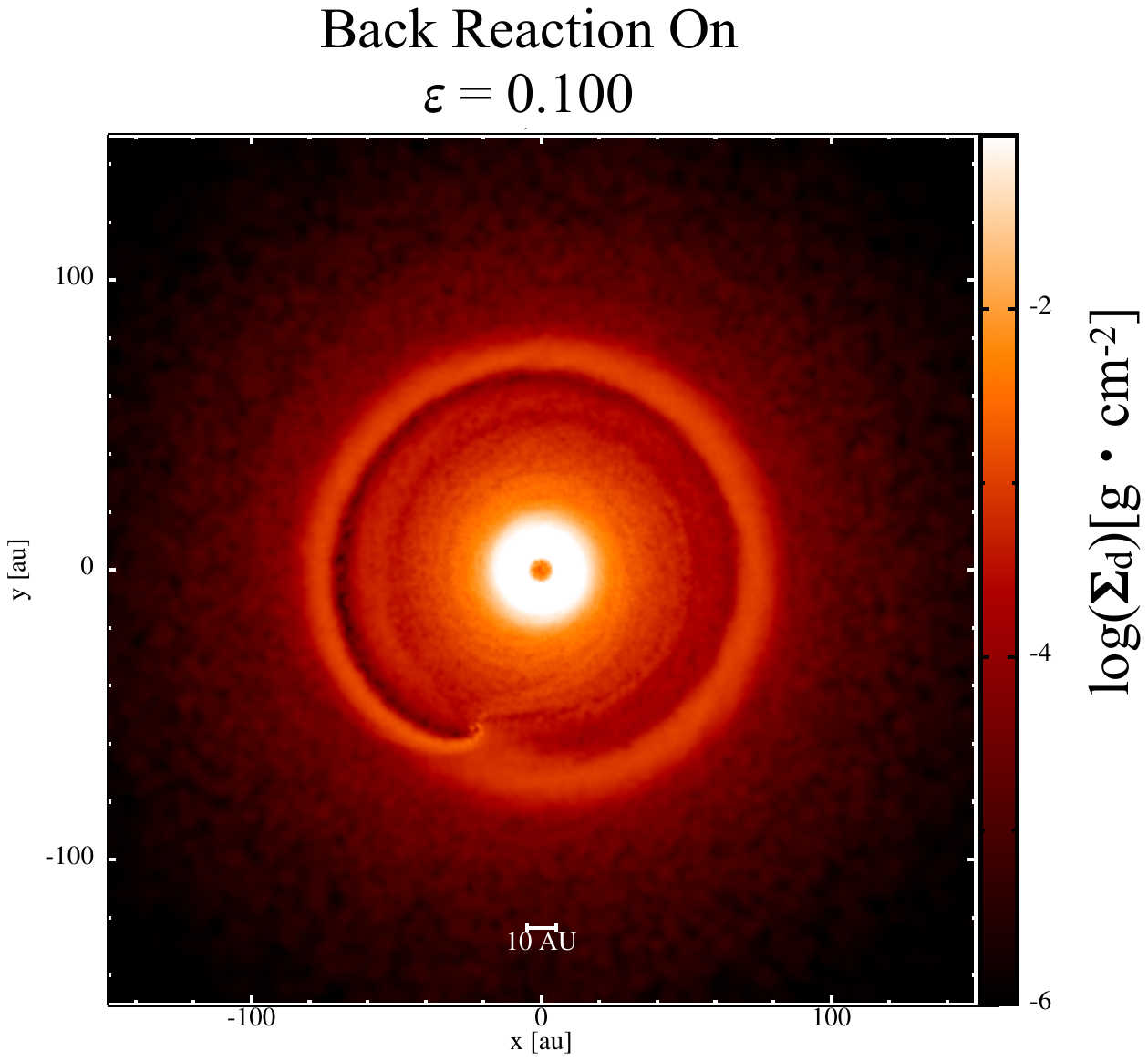} & \includegraphics[width=0.5\textwidth]{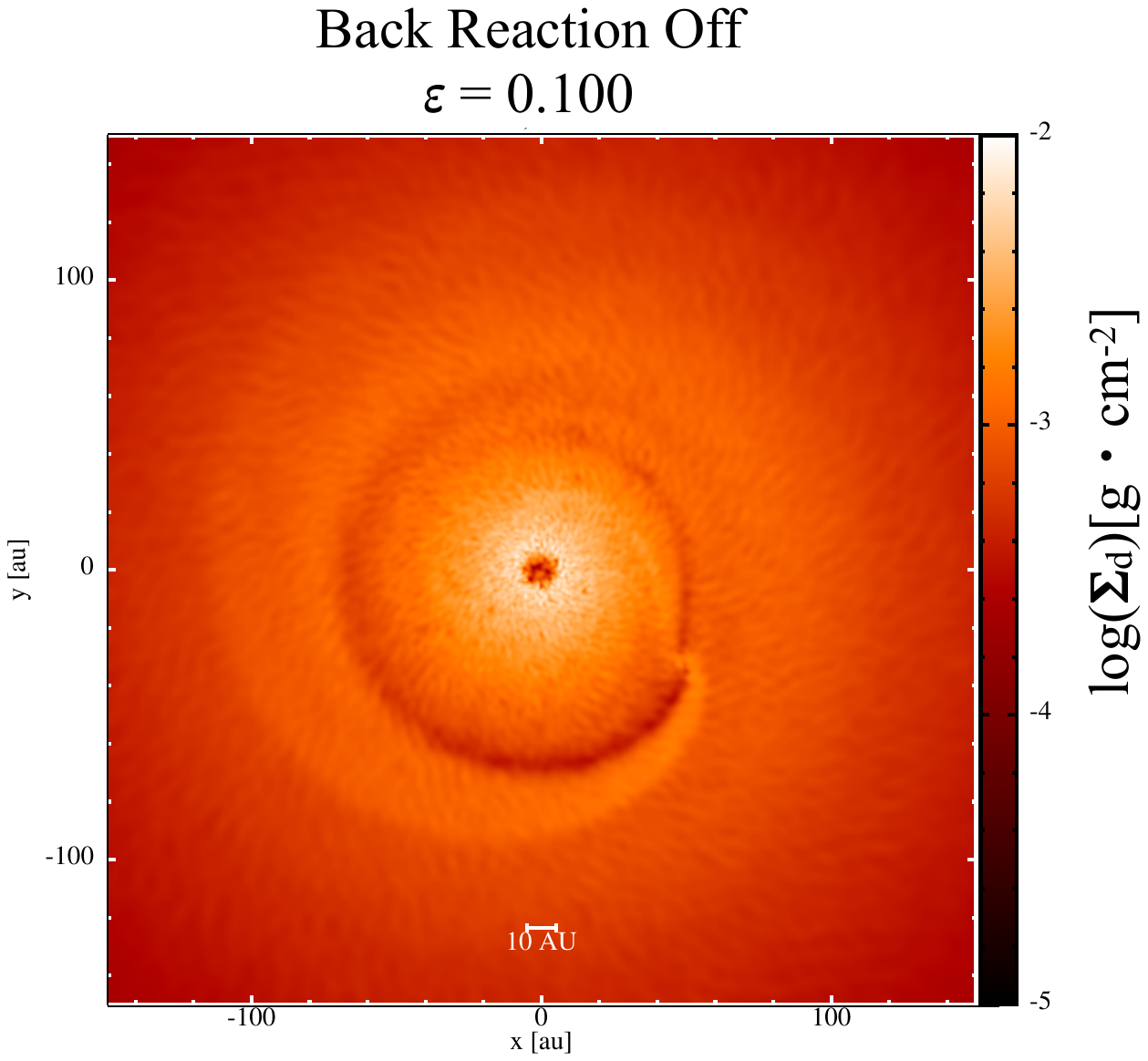} \\
         & \includegraphics[width=0.5\textwidth]{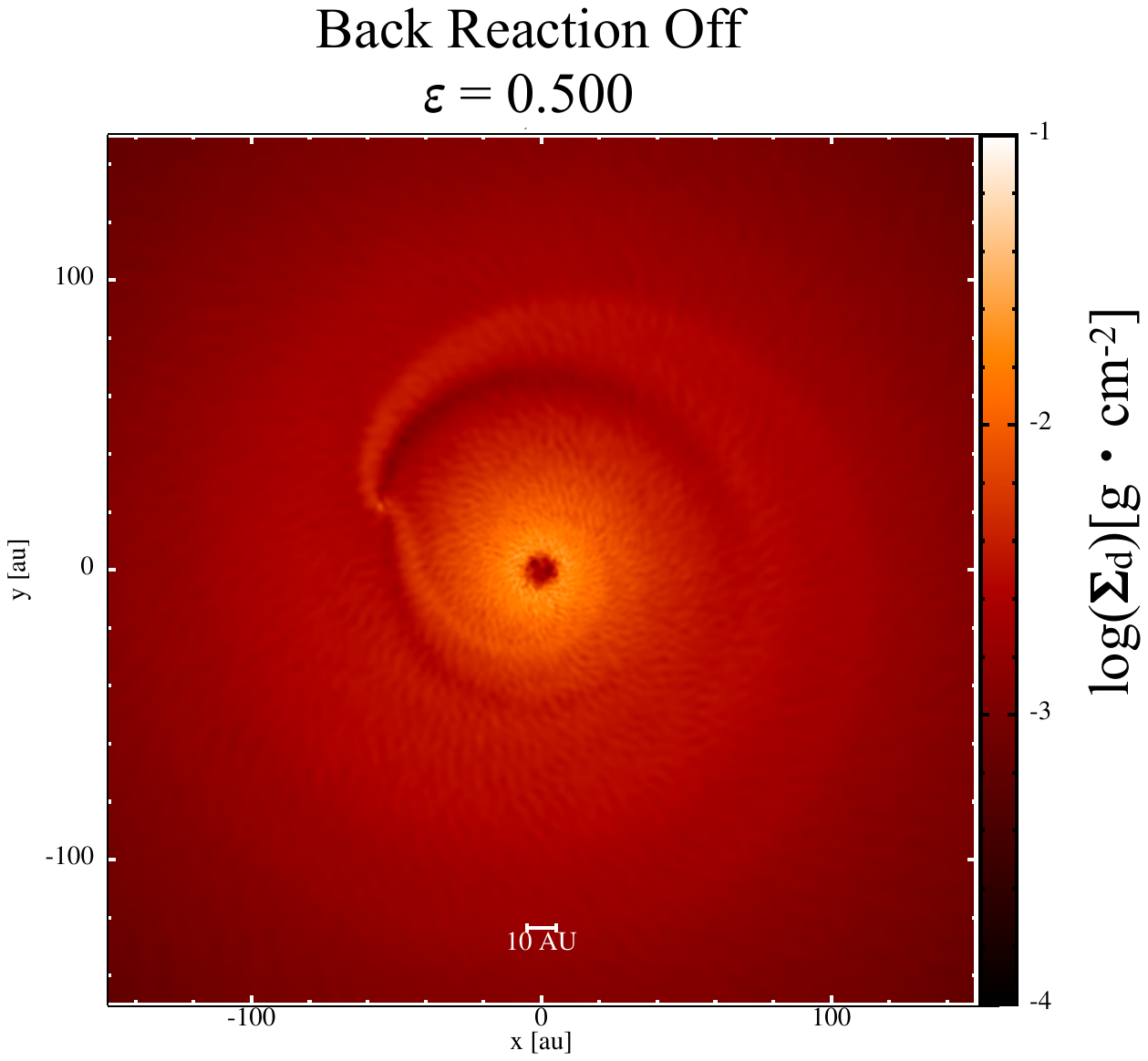} \\
    \end{tabular}
   \caption{Dust surface density of simulated discs with $a=0.1$mm. The left column is simulated with the back reaction, and the right column is simulated without the back reaction. The top row is $\epsilon = 0.10$ and the bottom row is $\epsilon = 0.50$. The simulated disc with $\mathrm{St}=0.50$ and $\epsilon = 0.50$ is not included due to simulation not reaching our minimum of 80 orbits before timesteps became prohibitive.}
    \label{fig:StokesDen2}
\end{figure*}

\bsp	
\label{lastpage}
\end{document}